\documentclass[fleqn,usenatbib]{mnras}

\usepackage[T1]{fontenc}
\usepackage{ae,aecompl}
 
\usepackage{graphicx}	
\usepackage{amsmath}	

\usepackage{amssymb}	

\title[{\it Chandra} spectra of 4U~1630-47]{{\it Chandra} high-resolution spectra of 4U~1630-47: the disappearance of the wind   }

\author[Gatuzz et al.]{
E. Gatuzz$^{1,2}$\thanks{E-mail: egatuzzs@eso.org},
M. D\'iaz Trigo$^{1}$,
J.C.A. Miller-Jones$^{3}$
and S. Migliari$^{4,5}$
\\
$^{1}$ESO, Karl-Schwarzschild-Strasse 2, D-85748 Garching bei M\"unchen, Germany\\
$^{2}$Excellence Cluster Universe, Boltzmannstr. 2, D-85748, Garching, Germany\\
$^{3}$International Centre for Radio Astronomy Research, Curtin University, G.P.O. Box U1987, Perth, WA, 6845, Australia\\
$^{4}$XMM-Newton Science Operations Centre, ESAC/ESA, Camino Bajo del Castillo s/n, Urb. Villafranca del Castillo, \\
28691 Villanueva de la Ca\~nada, Madrid, Spain\\
 $^{5}$Institute of Cosmos Sciences, University of Barcelona, Mart\'i i Franqu\`es 1, 08028 Barcelona, Spain\\
} 
\date{Accepted XXX. Received YYY; in original form ZZZ} 
\pubyear{2018} 
\begin{document}
 \label{firstpage}
\pagerange{\pageref{firstpage}--\pageref{lastpage}}
\maketitle 
\begin{abstract}
 We present the analysis of six {\it Chandra} X-ray high-resolution observations of the black hole low-mass X-ray binary 4U~1630-47 taken during its 2012-2013 outburst. {\rm Fe}~{\sc XXVI} K$\alpha$, K$\beta$, {\rm Fe}~{\sc XXV} K$\alpha$, K$\beta$ and {\rm Ca}~{\sc XX} K$\alpha$ blueshifted absorption lines were identified in the first four observations, which correspond to soft accretion states. The remaining observations, associated to intermediate and possibly hard accretion states, do not show significant absorption features down to equivalent width of 1 eV for both {\rm Fe}~{\sc XXVI} and {\rm Fe}~{\sc XXV}.  We inferred wind launching radii between  $1.2- 2.0$ ($10^{12}$ cm$/n$)$ \times 10^{11}$~cm and column densities $N({\rm H})> 10^{23}$ cm$^{-2}$. In the first four observations we found that thermal pressure is likely to be the dominant launching mechanism for the wind, although such conclusions depend on the assumed density. We used the spectral energy distributions obtained from our continuum modeling to compute thermal stability curves for all observations using the {\sc xstar} photoionization code. We found that the absence of lines in the transitional state cannot be attributed to an evolution of the plasma caused by thermal instabilities derived from the change in the continuum spectrum. In contrast, the disappearance of the wind could indicate an acceleration of the flow or that the plasma has been exhausted during the soft state.
\end{abstract}

\section{Introduction}\label{sec_in} 
In the last decade, analysis of black hole low-mass X-ray binaries (BH LMXBs) using X-ray spectra has shown the presence of photoionised plasmas in such systems \cite[see][and references therein]{dia16}. The plasmas can be found as a bound atmosphere or they can flow outwards with velocity blueshifts well above 1000 km/s \citep{kal09}. Although it is not clear which mechanism is responsible for the wind launching, the best candidates include thermal pressure, radiative pressure and magnetic pressure. Thermal pressure consists of the heating of the gas by the central X-ray source, producing an outflow at large distances from the central object, when the thermal velocity is larger than the local escape velocity \citep{beg83,woo96,net06,hig15,hig17}. Radiative pressure can be produced by electrons scattering, at near- or super-Eddington luminosities, or lines  \citep{roz14,has15,shi16}. However, it has been shown that, for BH LMXBs, radiation pressure cannot launch an outflow due to the low number of soft X-ray and UV lines \citep{pro02}. Finally, magnetic pressure or magnetocentrifugal forces can produce winds at small radii, although more work needs to be done, from the theoretical point of view, in order to reproduce the observed spectra  \citep{bla82,pro03,cha16,li14}. 

From the observational point of view, thermal winds have been favored by a majority of observations  \citep{kub07,nei12,roz14,dia16,all18,don18,tom18} although radiation pressure due to electrons \citep{kot00,kub07,nei09b,dia14,roz14,mil16d,shi16}  and magnetic forces \citep{mil06a,mil08,luk10,kin14,mil06a,mil16c,fuk17,tet18} have also been invoked to explain a handful of cases.

An open issue is the connection between the outflowing winds and the accretion state. During outburst, BH X-ray binaries show a hysteresis pattern in the hardness-intensity diagram that has been associated to transitions through different accretion states \citep{fen04,fen12}. Because the winds have been observed in a number of BHs  to be stronger in the soft accretion state in which jets are quenched \citep{mil06a,mil06dd,dia07,kub07,ued09bb,dia14} it was proposed that jets and winds were preventing each other from forming \citep{neil09}. However, it has been recently shown that disc winds and jets may co-exist. For example, \citet{hom16} found indications that some sources could produce winds and jets in the same accretion state (albeit with non-simultaneous observations) and concluded that if the LMXB luminosity is above a few tens of percent of the Eddington luminosity, disc winds and jets may co-exist. In the case of the BH LMXB V404 Cygni, an optical wind has been identified simultaneously with a radio jet by \citet{mun17}. Also, \citet{rah14} found a broad Pa$\beta$ absorption feature in the hard state of BH LMXB GX~339--4 using observations taken with the ESO/Very Large Telescope (VLT) that they attributed to a wind. Despite the efforts made to improve our understanding of these phenomena, multiple questions remain, including: What dictates the balance of power and the matter/radiation content of the disc, wind and jet? Is the disc-jet connection defined by the accretion flow only or does it depend on the compact object? And  how do winds affect the accretion process?  
 
The BH LMXB~4U~1630-47 constitutes an excellent laboratory to study the disc-jet connection. It has been identified as a recurrent transient \citep{jon76, par95,kuu97,tom05} with an inclination of $\sim$ 60--75 $^{\circ}$ \citep{kuu98,tom98}. Radio emission has been detected at flux levels always $<$ 3\,mJy\,beam$^{-1}$ and has been identified with the presence of jets in this system (\citealt{hje99}; \citealt{dia13}; but see \citealt{nei14} for a different interpretation). 
 
 {\it Suzaku} spectra of this source were analyzed by \citet{kub07} who first identified a highly ionized disc wind traced by {\rm Fe}~{\sc XXVI} and {\rm Fe}~{\sc XXV} absorption lines during the 2006 outburst. They concluded that thermal and radiative pressure processes can be part of the launching mechanism without discarding completely magnetic processes. \citet{dia14} analyzed {\it XMM-Newton} X-ray spectra obtained during the 2011-2013 outburst. They identified {\rm Fe}~{\sc XXVI}, {\rm Fe}~{\sc XXV}, {\rm Ni}~{\sc XXVIII}, {\rm Ni}~{\sc XXVII} and {\rm S}~{\sc XVI} absorption lines associated with a disc wind being thermally-radiatively driven during a soft state of the source. They followed the source across the transition from a soft state to a very high state and attributed the disappearance of the wind in a very high state to strong ionization of the wind due to the hardening of the spectrum and the increase of luminosity during that state. \citet{nei14} analyzed {\it Chandra} high-resolution spectra obtained in 2012 during the same outburst. They fitted the continuum of the observation taken on January 2012 with a disc blackbody and classified the source as being in a soft accretion state, with the presence of absorption lines due to {\rm Fe}~{\sc XXVI}, {\rm Fe}~{\sc XXV}, {\rm Ca}~{\sc XX}, {\rm Ni}~{\sc XXVIII} and {\rm Ni}~{\sc XXVII}. In contrast, they found that the continuum  of the observation taken in June 2012 shows an additional power-law component and no absorption lines.
  
With the aim of studying further the connection of the wind, jet and accretion state across state transitions, we obtained two simultaneous {\it Chandra} and {\it VLA} observations during a soft-to-hard state transition. For comparison, we also used four Chandra observations from an earlier time in the outburst that show significant line absorption \citep{nei14}.  The outline of this paper is as follows. In Section~\ref{sec_dat}, we describe the data selection and the reduction. The continuum modeling and the fit of the absorption lines identified are described in Sections~\ref{sec_cont} and \ref{sec_lines}, respectively. Results obtained by using photoionization models are reviewed in Section~\ref{sec_dis}. An analysis of the thermal stability curves derived for all observations is included in Section~\ref{sec_ther}. In Section~\ref{sec_dis2}, we discuss the possible launching mechanisms present in this system and Section~\ref{sec_con} summarizes the main results of our analysis. We assume a value of 10~kpc throughout this paper, as previous authors \citep{abe05,tom05,dia14} but note that \citet{kal18} have recently reported two potential distances of $4.7\pm0.3$ kpc and $11.5\pm 0.3$ kpc to the source based on an analysis of the dust scattering halo around the source.

 \begin{table*} 
\small
\caption{\label{tab_data}{\it Chandra} High-Energy Grating observations of 4U~1630-47.}
\centering 
\begin{tabular}{cccccccc}   
\hline
Label&ObsID & Date &   MJD & Exposure & Count-rate (cts/s) \\
&&&Start-time&(ks)&(1.5--10 keV)\\
\hline
Obs1&13714&	17 Jan 2012& 55943.2&	28.9	 & 14.8 \\
Obs2&13715&	20 Jan 2012& 55943.1&	29.2	 & 14.4 \\
Obs3&13716&	26 Jan 2012& 55943.1&	29.2	 & 13.7  \\
Obs4&13717&	30 Jan 2012& 55956.3&	29.4	 & 15.6  \\
Obs5&14441&	03 Jun 2012& 56081.9&	19.0	 & 20.8 \\
Obs6&15511&	25 Apr 2013& 56407.2&	49.4	 & 9.8   \\
Obs7&15524&	27 May 2013& 56439.7&	48.9	 & 0.6  \\
\end{tabular}
\end{table*}

   \begin{figure*}
        \begin{center}
\includegraphics[scale=0.38]{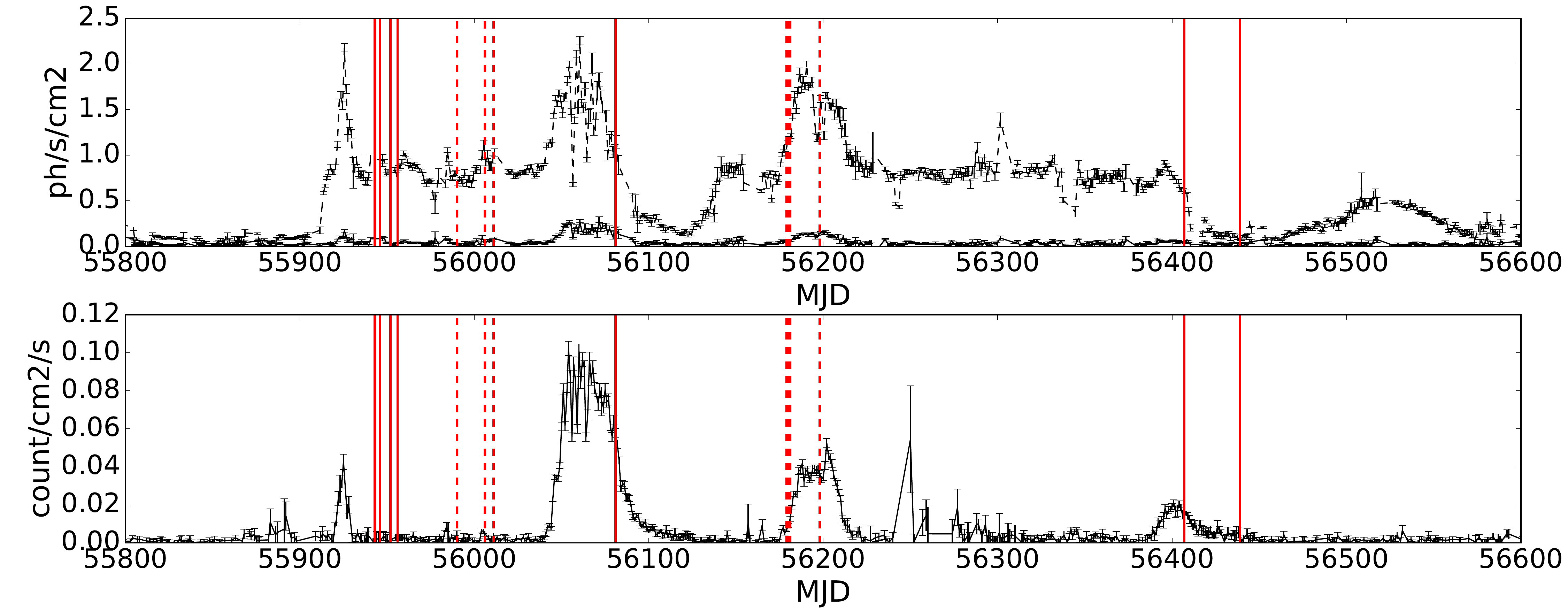}
      \caption{Top panel: {\it MAXI/ASM} daily average lightcurves of 4U~1630-47.  The black dashed line corresponds to the 2--20 keV lightcurve while black solid line corresponds to the 10--20 keV light curve. Bottom panel: {\it Swift/BAT} daily average lightcurve of the LMXB~4U~1630-47 in the 15--50 keV  energy range. In both panels vertical red solid lines indicate the {\it Chandra} observation dates while vertical red dashed lines indicate the {\it XMM-Newton} observations analyzed by \citet{dia14}.   }\label{fig_lc}
      \end{center}
   \end{figure*}
   
   \begin{figure} 
        \begin{center}
\includegraphics[scale=0.53]{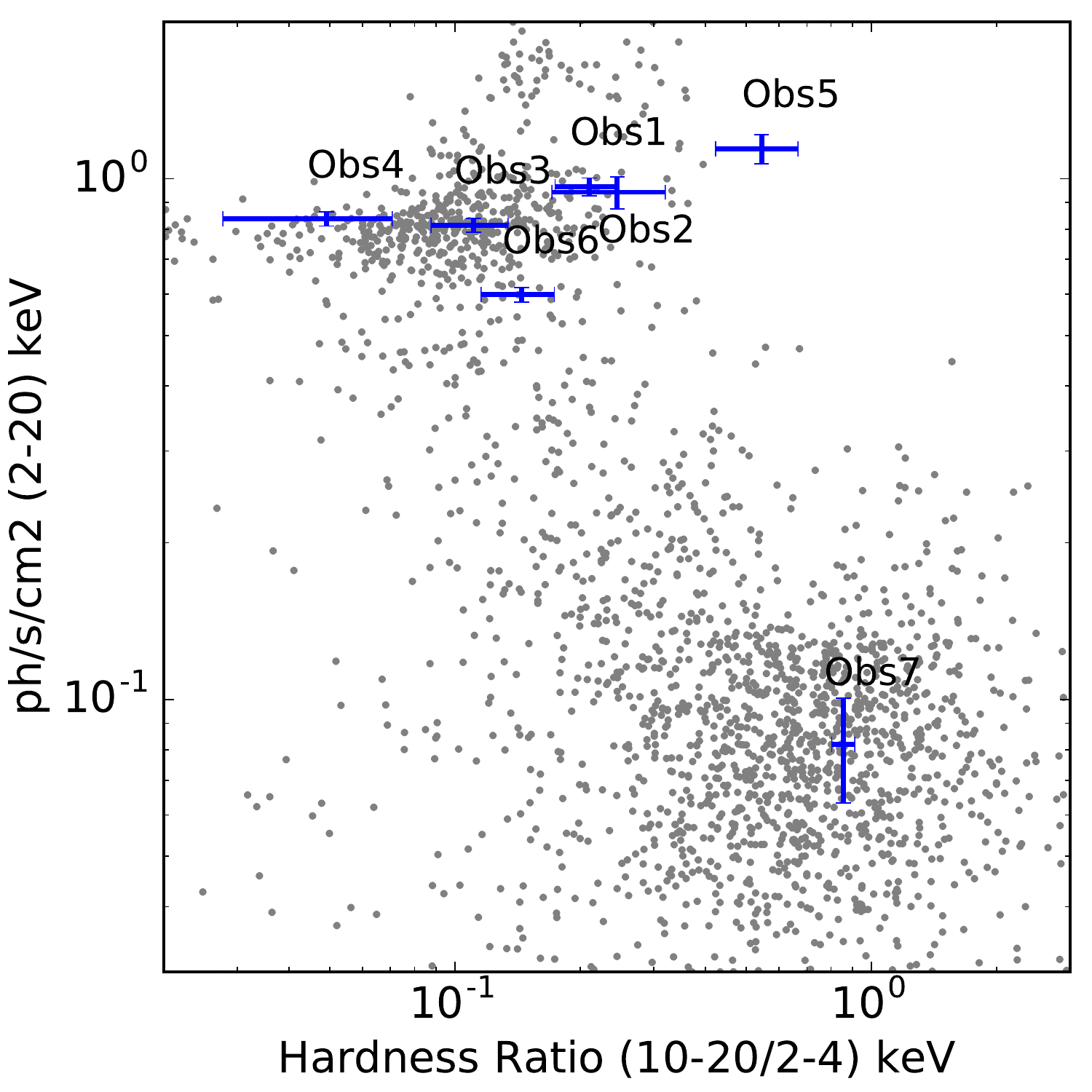}
      \caption{Hardness-intensity diagram of the 4U~1630-47 using {\it MAXI/ASM} daily average lightcurves. The positions of the pointed {\it Chandra} observations during the outburst are marked by blue crosses. }\label{fig_hr}
      \end{center}
   \end{figure}    
   
\begin{table*}
\scriptsize 
\caption{\label{tab_con}4U~1630-47 {\it Chandra} HEG  best-fit results. }
\centering
\begin{tabular}{llcccccc}
\\
Component&Parameter&Obs1&Obs2&Obs3&Obs4 &Obs6&Obs7\\
\hline
\hline
\\ 
\multicolumn{8}{c}{Model B: {\tt tbabs*(diskbb)}}\\
{\tt Tbabs} & $N({\rm H})$&$9.15\pm 0.05  $&$9.14\pm 0.05 $&$9.12\pm 0.05  $&$9.23 \pm 0.05 $ &$- $&$-$ \\
{\tt diskbb}&$kT_{in}$ &$1.54\pm 0.01  $&$1.49\pm 0.01 $&$1.52\pm 0.01  $&$ 1.58\pm 0.01 $ &$-  $&$- $ \\
& norm$_{dbb}$ &$113 \pm 3  $&$126\pm 3 $&$112\pm 3  $&$107 \pm 3 $&$-  $&$-  $  \\
Statistic&$\chi^{2}$/d.of.&$3056 /2810 $&$2968/2810 $&$ 2974/2810 $&$ 3084/2810 $&$ - $&$- $  \\
&red-$\chi^{2}$&$1.08 $ &$1.05 $ &$1.05 $ &$1.09 $ &$- $ &$-$   \\
Count-rate&Model &6.0$\times 10^{-5}$   &4.9$\times 10^{-5}$   &5.3$\times 10^{-5}$   & 7.9$\times 10^{-5}$  &$-$  &$-$   \\
(15-50 keV)&{\it Swift/BAT} & $<$ 5.2$\times 10^{-4}$ & $<$ 9.5$\times 10^{-4}$ & $<$ 1.1$\times 10^{-3}$   &   $<$ 6.4$\times 10^{-4}$  &$-$  & $-$   \\
Flux&(0.0136--13.60 keV)&1.4$\times 10^{-8}$ &1.4$\times 10^{-8}$  &1.3$\times 10^{-8}$ & 1.5$\times 10^{-8}$ &$-$ &$-$    \\
&(1.5--10 keV)& 9.9$\times 10^{-9}$ & 9.7$\times 10^{-9}$  & 9.3$\times 10^{-9}$ &  1.0$\times 10^{-8}$ &$-$ &$-$    \\
&(15--50 keV)& 2.3$\times 10^{-11}$ & 2.6$\times 10^{-10}$  & 1.9$\times 10^{-11}$ & 3.1$\times 10^{-11}$ &$-$ &$-$    \\
\hline
 \\ 
\multicolumn{8}{c}{Model C: {\tt tbabs*(powerlaw+diskbb)}}\\
{\tt Tbabs} & $N({\rm H})$ &$9.15 \pm 0.04 $&$9.14 \pm 0.04 $&$9.12 \pm 0.04   $&$9.22 \pm 0.05  $  &$9.41\pm 0.27  $&$ 9.10$ (fixed)\\
{\tt powerlaw}&$\Gamma$ &$2.5$ (fixed)&$2.5$ (fixed)&$2.5$ (fixed)&$2.5$ (fixed) &$2.4\pm 0.2  $&$2.1\pm 0.3 $  \\
& norm$_{pow}$ &$ < 0.08   $&$<0.07  $&$ <0.07   $&$ <0.08  $& $ 1.9 _{-0.9}^{+1.5} $&$0.10\pm 0.05 $ \\
{\tt diskbb}&$kT_{in}$  &$1.54 \pm 0.01 $&$1.49 \pm 0.01  $&$1.52 \pm 0.01   $&$1.58 \pm 0.01  $ &$1.21\pm 0.03  $&$0.60\pm 0.03 $ \\
&norm$_{dbb}$ &$113 \pm 3 $&$126\pm 3 $&$112 \pm 3   $&$107 \pm 3  $    &$138\pm 23  $&$284_{-55}^{+75} $  \\
Statistic&$\chi^{2}$/d.of.&$ 3056/2810 $&$2968/2810 $&$2974/2810 $&$ 3084/2810 $& $ 2974/2810 $&$ 2490/2810 $ \\
&red-$\chi^{2}$&$1.08$ &$1.05$ &$1.06$ &$1.09$   &$1.06 $ &$0.88 $  \\
Count-rate&Model &6.5$\times 10^{-5}$ &5.0$\times 10^{-5}$&5.4$\times 10^{-5}$&8.4$\times 10^{-5}$   & 2.7$\times 10^{-3}$ &  3.8$\times 10^{-4}$ \\
(15-50 keV)&{\it Swift/BAT} &  $<$ 5.2$\times 10^{-4}$ & $<$ 9.5$\times 10^{-4}$ & $<$ 1.1$\times 10^{-3}$   &   $<$ 6.4$\times 10^{-4}$  & (2.5  $\pm$ 0.2)$\times 10^{-3}$  &(4.1  $\pm$ 3.1)$\times 10^{-4}$ \\
Flux&(0.0136--13.60 keV)&1.4$\times 10^{-8}$ &1.4$\times 10^{-8}$ &1.3$\times 10^{-8}$ & 1.4$\times 10^{-8}$ &  4.9$\times 10^{-8}$ & 2.1$\times 10^{-9}$  \\
&(1.5--10 keV)& 1.9$\times 10^{-9}$ & 9.7$\times 10^{-9}$  & 1.0$\times 10^{-8}$ &  1.0$\times 10^{-8}$ &7.9$\times 10^{-9}$ &5.4$\times 10^{-10}$    \\
&(15--50 keV)& 2.3$\times 10^{-11}$ & 2.5$\times 10^{-10}$  & 1.7$\times 10^{-11}$ &  2.9$\times 10^{-11}$ &1.0$\times 10^{-9}$ &1.4$\times 10^{-10}$    \\
 \hline
  \\ 
 \multicolumn{8}{c}{Model D: {\tt tbabs*simpl(diskbb)}}\\
{\tt Tbabs} & $N({\rm H})$ &$9.15 \pm 0.01   $&$9.13\pm 0.02 $&$9.12\pm 0.03   $&$ 9.23\pm 0.05 $ &$9.03\pm 0.09  $&$ 9.10$ (fixed)\\
{\tt simpl}&$\Gamma$ &$<2.00 $&$<2.00 $&$ <2.00  $&$<2.00 $ &$3.7 \pm 0.3 $&$2.1\pm 0.3 $ \\
&FracSca &$< 0.01  $&$< 0.01 $&$< 0.01   $&$< 0.01$ &$0.8_{-0.3}^{+0.8}  $&$0.18\pm 0.03 $\\
{\tt diskbb}&$kT_{in}$ &$1.53\pm 0.01  $&$1.49\pm 0.01 $&$ 1.52\pm 0.01 $&$ 1.58\pm 0.01 $ &$0.90\pm 0.01  $&$0.58\pm 0.04 $ \\
&norm$_{dbb}$ &$115_{-4}^{+2E06}$&$125_{-3}^{+59}$&$112_{-1}^{+105}   $&$ 108\pm 4 $ &$560_{-1}^{+350}  $&$431^{+181}_{-113} $ \\
Statistic&$\chi^{2}$/d.of.&$ 3056/2810  $&$ 2971/2810 $&$2974/2810$&$ 3085/2810 $ &$2958/2810 $&$ 2493/2810 $ \\
&red-$\chi^{2}$&$1.08 $ &$1.05 $ &$1.06 $ &$1.09 $   &$1.05 $ &$0.88$  \\
Count-rate&Model&5.4$\times 10^{-5}$  & 4.4$\times 10^{-5}$  &5.4$\times 10^{-5}$ &7.4$\times 10^{-5}$     &2.2$\times 10^{-3}$  & 5.0$\times 10^{-4}$ \\
(15-50 keV)&{\it Swift/BAT} & $<$ 5.2$\times 10^{-4}$ & $<$ 9.5$\times 10^{-4}$ & $<$ 1.1$\times 10^{-3}$   &   $<$ 6.4$\times 10^{-4}$       & (2.5 $\pm$ 0.2)$\times 10^{-3}$  &(4.1 $\pm$ 3.1)$\times 10^{-4}$ \\
Flux&(0.0136--13.60 keV)&1.4$\times 10^{-8}$ &1.4$\times 10^{-8}$ &1.3$\times 10^{-8}$ & 1.4$\times 10^{-8}$   &1.1$\times 10^{-8}$  &1.4$\times 10^{-9}$  \\
&(1.5--10 keV)& 9.9$\times 10^{-9}$ & 9.7$\times 10^{-9}$  & 9.3$\times 10^{-9}$ &  1.0$\times 10^{-8}$ &7.4$\times 10^{-9}$ & 5.5$\times 10^{-10}$   \\
&(15--50 keV)& 3.0$\times 10^{-10}$ & 1.7$\times 10^{-11}$  & 1.9$\times 10^{-11}$ &  2.9$\times 10^{-11}$ &3.4$\times 10^{-10}$ & 4.4$\times 10^{-10}$    \\
\hline
\\
\multicolumn{8}{l}{Hydrogen column density ``$N({\rm H})$'' in  units of $\times 10^{22}$~cm$^{-2}$. Temperature at inner disc radius ``$kT_{in}$'' in units of keV.   }\\
\multicolumn{8}{l}{{\tt Powerlaw} normalization ``norm$_{pow}$'' in units of ph/keV/cm$^{2}$/s at 1 keV.  }\\
\multicolumn{8}{l}{{\tt diskbb} normalization ``norm$_{dbb}=(R_{in}/D_{10})^{2}\cos\theta$'' where $R_{in}$ is the inner disc radius,  }\\
\multicolumn{8}{l}{ $D_{10}$ is the distance in units of 10~kpc and $\theta$ is the inclination of the disc. }\\
\multicolumn{8}{l}{Count-rate {\it Swift/BAT} are daily averaged count rates.}\\
\multicolumn{8}{l}{Count-rate model refers to the count rate predicted by the model for the {\it Swift/BAT} energy band.}\\
\multicolumn{8}{l}{Unabsorbed fluxes are given in units of ergs cm$^{-2}$ s$^{-1}$.}
\end{tabular}
\end{table*}
 
   \begin{figure*} 
        \begin{center}
\includegraphics[scale=0.36]{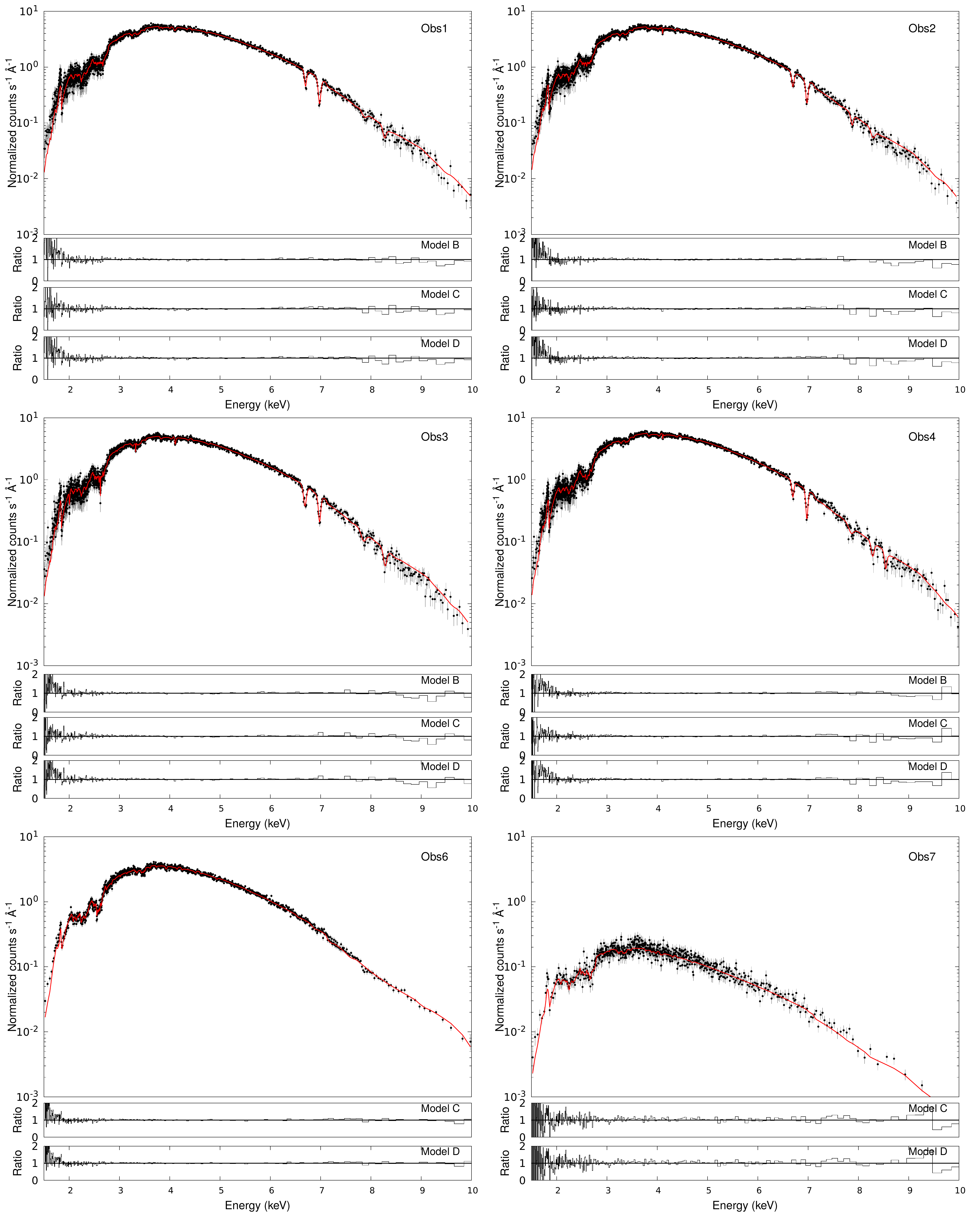}
      \caption{4U~1630-47 best continuum fit for all observations analyzed. Red solid lines indicate the best-fit obtained with model B (Obs~1-4) and model C (Obs~6-7).  Lower panels indicate the data/model ratios obtained for the different models described in Table~\ref{tab_con}. }\label{fig_dat_1}
      \end{center}
   \end{figure*}   
 
   \begin{figure} 
        \begin{center}
\includegraphics[scale=0.32]{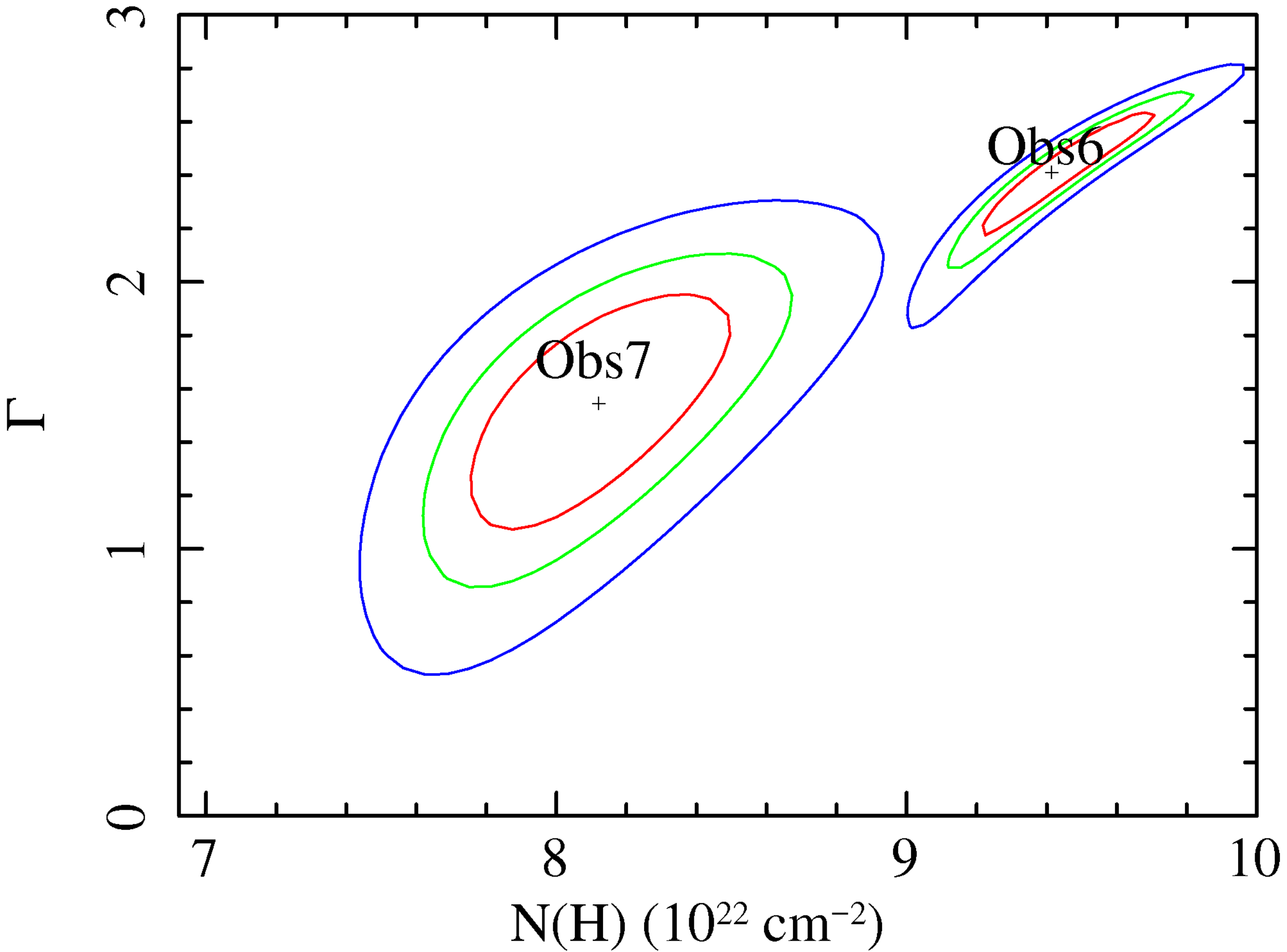}
      \caption{Contour plots of the $N({\rm H})$ and the $\Gamma$ parameters obtained from Model~C for Obs~6 and Obs~7. }\label{fig_contour}
      \end{center}
   \end{figure} 
   
\begin{table*}
\scriptsize 
\caption{\label{tab_gauss}4U~1630-47 {\it Chandra} HEG  Gaussian features included in the best-fit models listed in Table~\ref{tab_gauss}.}
\centering
\begin{tabular}{llccccccc} 
\\
Ion&Parameter&Obs1&Obs2&Obs3&Obs4 &Obs6&Obs7   \\
 
\hline
\hline
\\  
 {\rm Fe}~{\sc XXVI} K$\beta$&Energy  &  $8.27\pm 0.02  $&  $8.29 \pm 0.02   $&  $ 8.28 \pm 0.01   $&  $8.28 \pm 0.01  $ &  $8.28$(fixed) & $8.28$(fixed)\\
 &Wavelength  &$1.499\pm	0.004$  & $1.495\pm	0.004 $  &$ 1.497\pm	0.002 $ & $ 1.497\pm	0.002 $ &$1.49$(fixed)&$1.49$(fixed)\\
&$\sigma$&  $0.03\pm 0.02   $&  $0.04\pm 0.02  $&  $ 0.03\pm 0.01  $&  $<0.03  $& $0.03$(fixed)& $0.03$(fixed)\\
&norm&  $0.0007\pm 0.0002  $&  $0.0007\pm 0.0004  $&  $ 0.0009\pm 0.0002  $&  $0.0008\pm 0.0002 $&$<0.0001$&$<0.0001$\\
&EW  &  $36\pm 10   $&  $38\pm  2   $&  $ 49\pm 11  $&  $36\pm 9  $&$<1$&$<1$\\ 
{\rm Fe}~{\sc XXV} K$\beta$&Energy  &  $7.87 \pm 0.03  $&  $7.87\pm 0.01    $&  $ 7.85 \pm 0.02 $&  $7.86\pm 0.02  $ & $7.87$(fixed) & $7.87$(fixed)\\
 &Wavelength &$1.575\pm	0.006 $  & $1.575\pm 0.002 $   & $1.579\pm	0.004 $  & $1.577\pm 0.004 $&$1.57$(fixed)&$1.57$(fixed) \\
&$\sigma$&  $ 0.04 \pm 0.02 $&  $0.02 \pm 0.01  $&  $ 0.04 \pm 0.01   $&  $0.04 \pm 0.02 $&$0.03$(fixed)&$0.03$(fixed)\\
&norm&  $0.0006\pm 0.0002  $&  $0.0006\pm 0.0001  $&  $ 0.0008\pm 0.0002  $&  $0.0009\pm 0.0002 $&$<0.0003$&$<0.0001$\\
&EW  &  $25\pm 8   $&  $27\pm 5$&  $ 37 \pm 9    $&  $31 \pm 7  $&$<8$&$<1$\\
 {\rm Fe}~{\sc XXVI} K$\alpha$&Energy  &  $6.979 \pm 0.003    $&  $6.978\pm 0.003  $&  $ 6.976\pm 0.003 $& $6.975 \pm 0.002 $ & $6.97$(fixed) & $6.97$(fixed)\\
 &Wavelength  & $1.776\pm 0.001$ & $1.777\pm 0.001$   & $1.777\pm 0.001$  & $1.777\pm 0.001$&$1.77$(fixed)&$1.77$(fixed) \\
&$\sigma$&  $0.024 \pm 0.002   $&  $0.023\pm  0.003 $&  $ 0.023 \pm 0.003  $&  $0.019\pm 0.002  $&$0.022$(fixed)&$0.022$(fixed)\\
&norm&  $0.0022\pm  0.0001$&  $0.0019\pm 0.0001  $&  $ 0.0019\pm 0.0001  $&  $0.0022\pm 0.0001 $&$<0.0001$&$<0.0001$\\
&EW  &  $54 \pm 2   $&  $49\pm 3  $&  $ 53\pm 3   $&  $48\pm 2  $&$<1$&$<1$\\
 {\rm Fe}~{\sc XXV} K$\alpha$&Energy  &  $6.700 \pm 0.003  $&  $6.697 \pm 0.004 $&  $ 6.689\pm 0.003   $&  $6.697 \pm 0.003 $ & $6.7$(fixed) & $6.7$(fixed)\\
  &Wavelength  & $1.851\pm 0.001$ & $1.851\pm 0.001 $   & $1.854\pm	0.001$  & $1.851\pm 0.001$&$1.85$(fixed)&$1.85$(fixed) \\
&$\sigma$ &  $0.017 \pm 0.003  $&  $0.022\pm 0.003    $&  $ 0.029\pm 0.003  $&  $0.019\pm 0.003 $&$0.019$(fixed)&$0.019$(fixed)\\
&norm&  $0.0015\pm 0.0001 $&  $0.0015\pm 0.0001  $&  $ 0.0021\pm 0.0001  $&  $0.0016\pm 0.0001 $&$<0.0001$&$<0.0001$\\
&EW &  $31\pm 2    $&  $33 \pm 2  $&  $ 47 \pm 2    $&  $30 \pm 2  $&$<1$&$<1$\\
{\rm Ca}~{\sc XX} K$\alpha$&Energy &  $4.109 \pm 0.006   $&  $4.107 \pm 0.002   $&  $ 4.109 \pm 0.002   $&  $4.111\pm 0.003  $ &$-  $&$-  $\\
 &Wavelength  &$3.017\pm 0.004 $  & $3.019\pm	0.001 $   &$3.017\pm 0.001 $   &$3.016\pm 0.002 $ &$-  $&$-  $ \\
&$\sigma$&  $0.015 \pm 0.005   $&  $<0.003   $&  $ 0.010\pm 0.002  $&  $<0.001  $ &$-  $&$-  $\\
&norm&  $0.0005\pm 0.0001  $&  $0.0003\pm 0.0001  $&  $ 0.0007\pm 0.0001  $&  $0.0003\pm 0.0001 $&$-  $&$-  $\\
&EW &  $36.0\pm 0.7   $&  $24.0\pm  0.8  $&  $47.0\pm  0.6  $&  $23\pm 0.7  $&$-  $&$-  $\\
{\rm Ar}~{\sc XVIII} K$\alpha$&Energy &  $-  $&  $-  $&  $ 3.323\pm 0.001   $&  $- $&$-  $&$-  $\\
 &Wavelength  & $-  $ & $-  $   & $3.731\pm 0.001 $  &$-$&$-  $&$-  $  \\
&$\sigma$&  $-  $&  $-  $&  $ <0.002  $&  $- $&$-  $&$-  $\\
&norm&  $-  $&  $-  $&  $ 0.0003\pm 0.0001  $&  $- $&$-  $&$-  $\\
&EW &  $-  $&  $-  $&  $ 24\pm 0.8   $&  $- $&$-  $&$-  $\\ 
{\rm S}~{\sc XVI} K$\alpha$&Energy  &  $-  $&  $-  $&  $ 2.622 \pm 0.001   $&  $- $&$-  $&$-  $\\
 &Wavelength & $-  $ & $-  $   & $4.729\pm 0.002$  & $-$&$-  $&$-  $ \\
&$\sigma$&  $-  $&  $-  $&  $ 0.0032\pm 0.0009  $&  $- $&$-  $&$-  $\\
&norm&  $-  $&  $-  $&  $ 0.0003\pm 0.0001  $&  $- $&$-  $&$-  $\\
&EW &  $-  $&  $-  $&  $ 5\pm  2  $&  $- $&$-  $&$-  $\\  
\\
\\
\multicolumn{8}{l}{Energies and $\sigma$ are given in keV. Equivalent widths (EWs) are given in eV. Wavelengths are given in \AA.}\\
\multicolumn{8}{l}{Gaussian normalizations are given in photons cm$^{-2}$ s$^{-1}$.}
\end{tabular}
\end{table*}
 
 \begin{table*}
\small
\caption{\label{tab_warm}Best fits to the {\it Chandra} HEG spectra for Obs~1-4 using model B but substituting the Gaussian features by a warm absorber component.}
\centering
\begin{tabular}{llcccc}
\\
Component & Parameter&Obs $\#$1&Obs $\#$2&Obs $\#$3&Obs $\#$4 \\
\hline
\hline
\\  

\multicolumn{6}{c}{Model: {\tt tbabs*warmabs*(diskbb)}}\\
{\tt Tbabs} & $N({\rm H})$&$9.19\pm 0.07 $&$9.08\pm 0.06 $&$9.10 \pm 0.04 $&$9.16\pm 0.03 $ \\
{\tt diskbb}&$T_{in}$&$1.54\pm 0.01  $&$1.50\pm 0.01 $&$1.51\pm 0.01$&$1.59\pm 0.01 $ \\
&norm$_{dbb}$&$111\pm 1  $&$122\pm 2  $&$116 \pm 1  $&$103 \pm 1  $ \\
{\tt Warmabs} & $\log(N({\rm H})/10^{22})$&$1.17\pm 0.07 $&$1.29\pm 0.08 $&$1.08\pm 0.12 $&$1.30\pm 0.09 $ \\
& $\log(\xi)$&$3.99\pm 0.09 $&$4.00_{-0.14}^{+0.10} $&$3.55\pm 0.22 $&$3.92\pm 0.10 $ \\
& $v_{turb}$&$194\pm 9 $&$148_{-13}^{+6}  $&$200\pm 27 $&$157\pm 14 $ \\
& $z$&$ -(0.0020\pm 0.0001) $&$-(0.0019\pm 0.0002) $&$-(0.0010\pm 0.0002) $&$-(0.0019\pm 0.0001) $ \\
Flux&(0.0136--13.60 keV)&1.37$\times 10^{-8}$ &1.36$\times 10^{-8}$ &1.29$\times 10^{-8}$ & 1.44$\times 10^{-8}$     \\ 
Statistic &&$  2969/2803 $&$ 2840/2803 $&$ 2987/2803 $&$  2914/2803$\\
red-$\chi^{2}$&&$1.05$ &$1.01 $ &$ 1.06$ &$1.03 $ \\
\hline     
\\
\\
\multicolumn{6}{l}{ $N({\rm H})$ in  units of $\times 10^{22}$~cm$^{-2}$. $kT_{in}$ in units of keV.   }\\
\multicolumn{6}{l}{ norm$_{dbb}=(R_{in}/D_{10})^{2}\cos\theta$ where $R_{in}$ is the inner disc radius, $D_{10}$ is the distance in units of 10~kpc  } \\
\multicolumn{6}{l}{    and $\theta$ is the inclination of the disc. }
\end{tabular}
\end{table*}
 
   \begin{figure*} 
        \begin{center}
\includegraphics[scale=0.28]{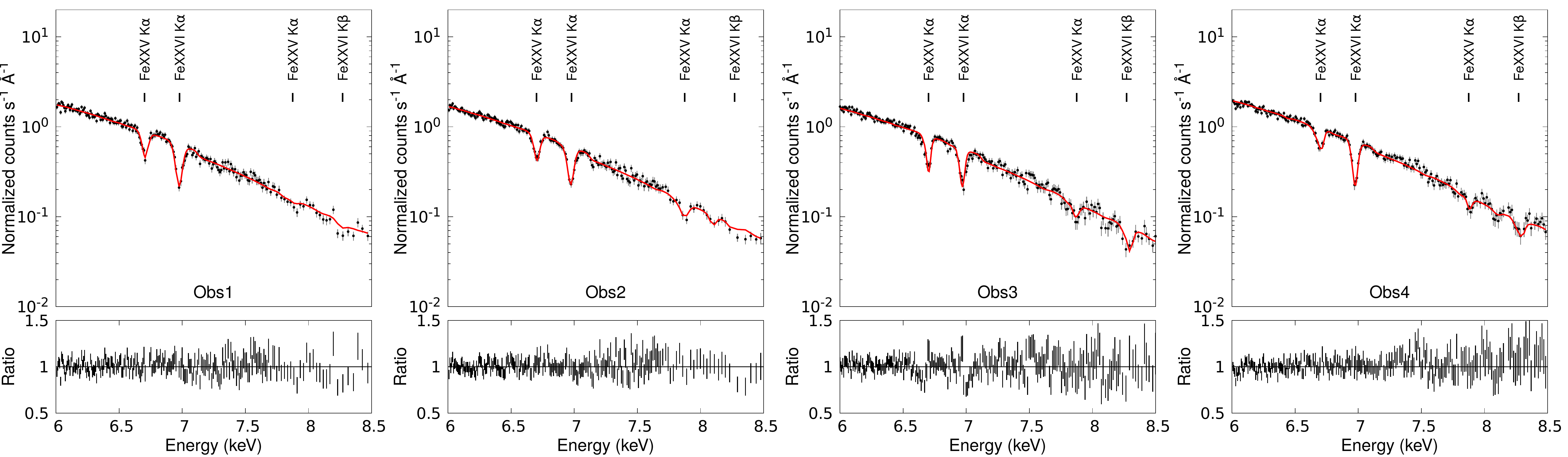}\\
\includegraphics[scale=0.28]{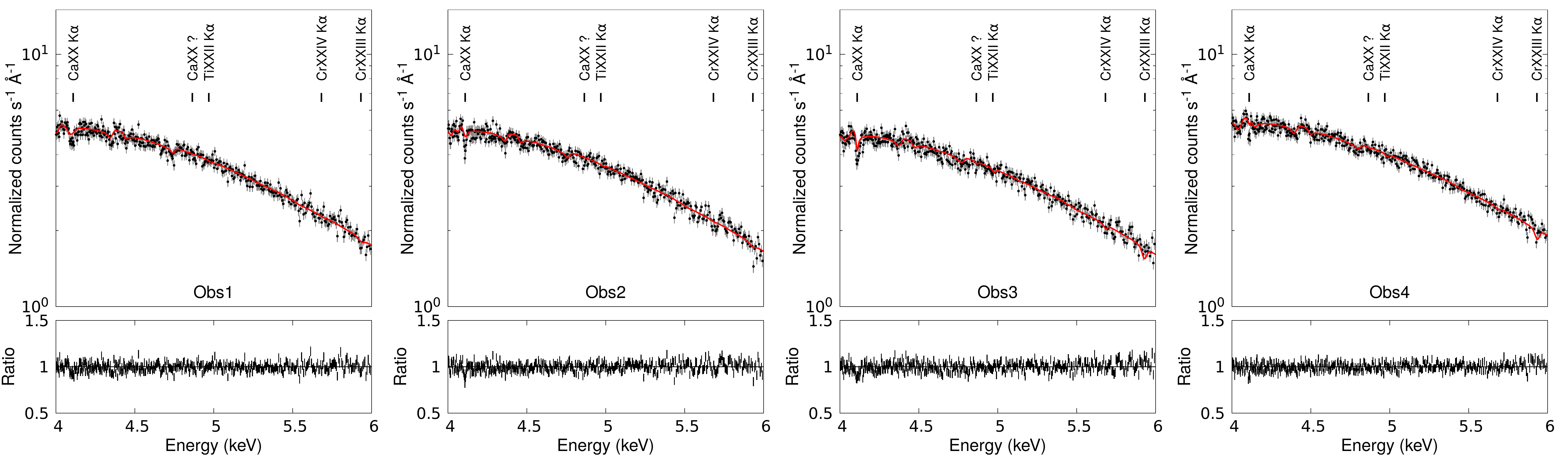}\\
\includegraphics[scale=0.28]{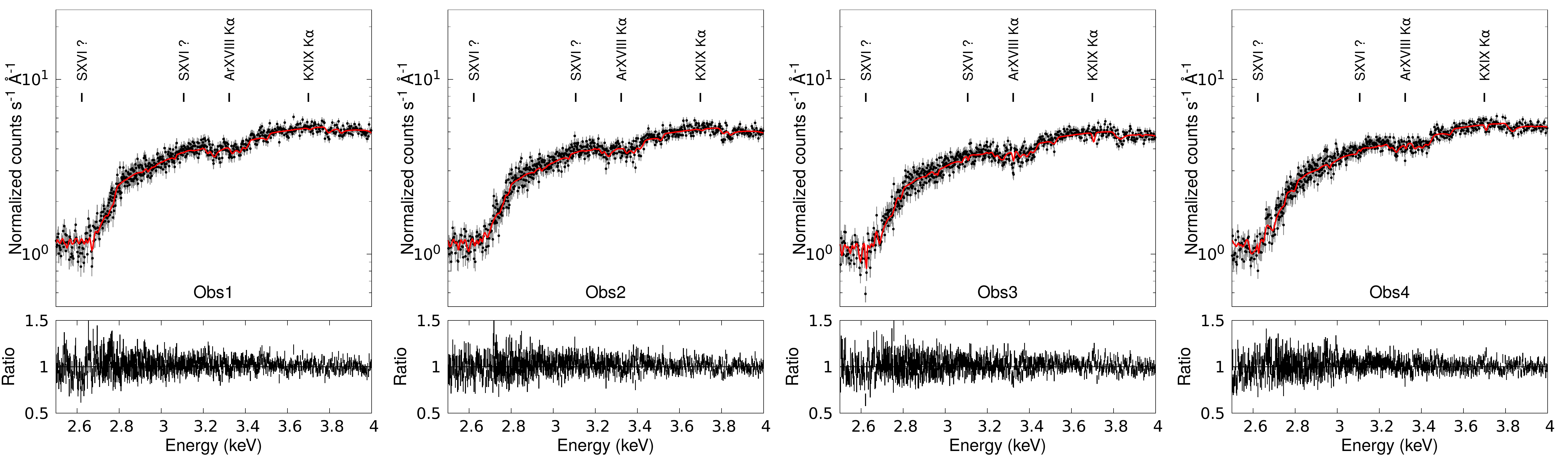}
      \caption{ Best fit results for Obs 1-4 modeled with {\tt warmabs} (See Section~\ref{tab_warm}). The main lines associated to the highly ionized absorber are indicated. The spectra have been rebinned for illustrative purposes.}\label{iron_warm}
      \end{center}
   \end{figure*}  
    
    \begin{figure} 
        \begin{center}
\includegraphics[scale=0.5]{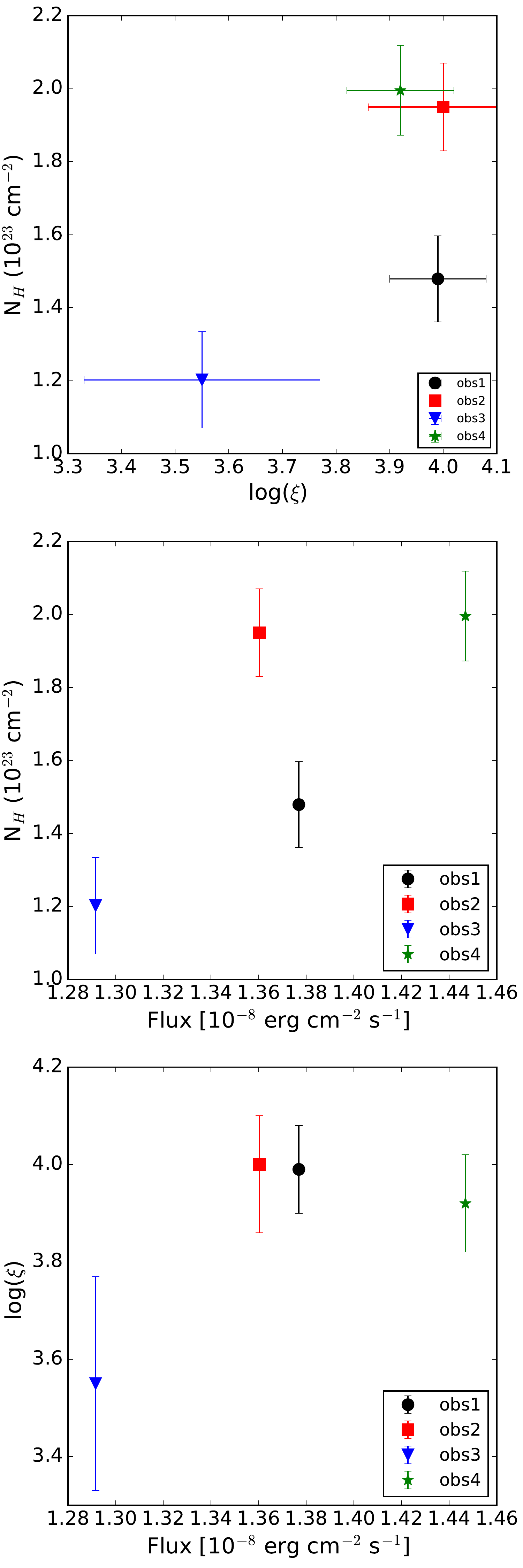}
      \caption{Upper panel: $N({\rm H})$ versus $\log{\xi}$ obtained from the {\tt warmabs} fit for Obs~1-4. Middle panel: $N({\rm H})$ versus unabsorbed flux in the 0.013 -13.6 keV energy range. Lower panel: $\log{\xi}$ versus the unabsorbed flux in the 0.013 -13.6 keV energy range.}\label{ng_logxi}
      \end{center}
   \end{figure} 
    
   \begin{figure} 
        \begin{center}
\includegraphics[scale=0.5]{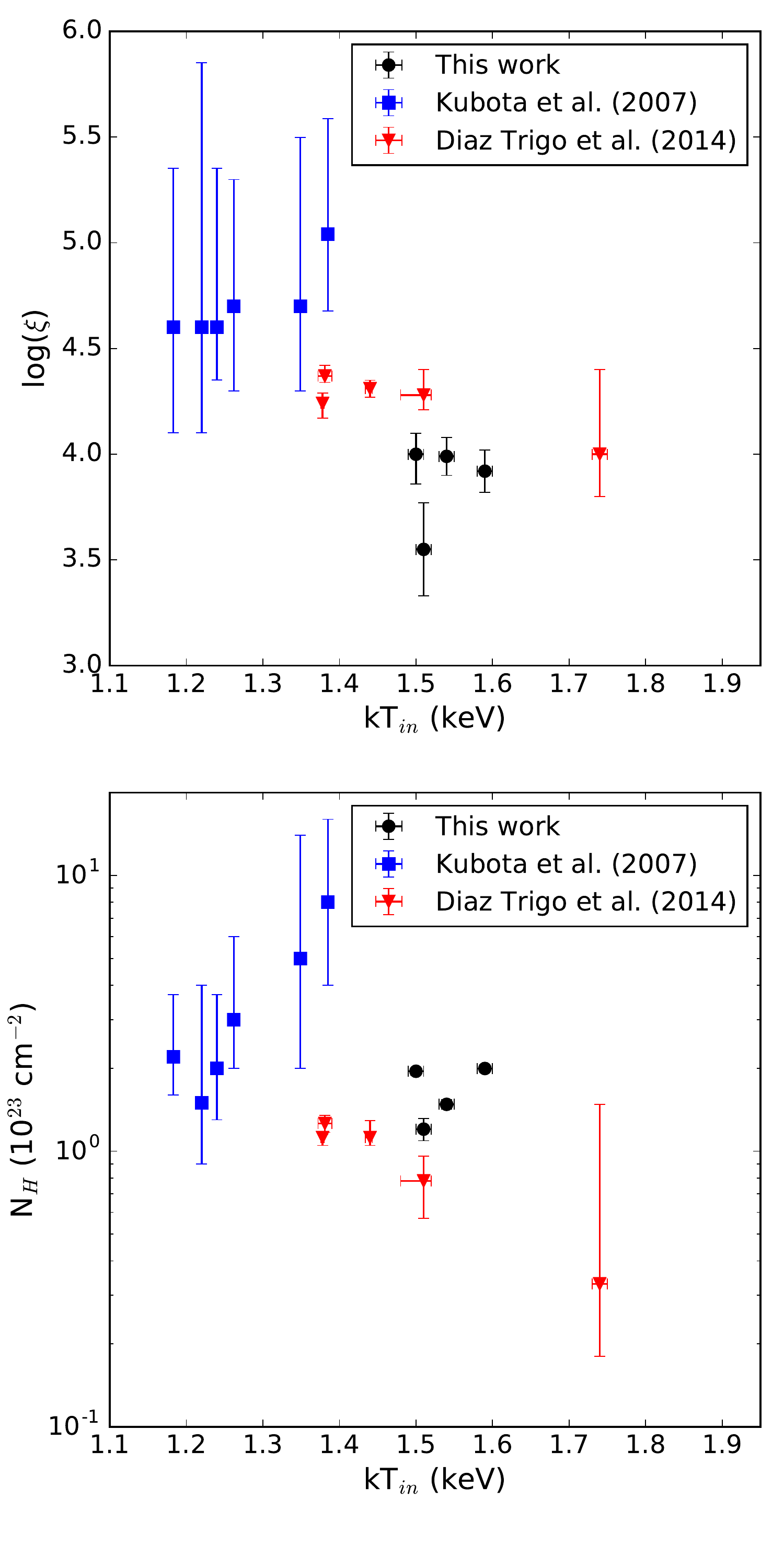}
      \caption{Upper panel: $\log{\xi}$ versus $kT_{in}$ obtained from the {\tt warmabs} fit for Obs~1-4 (Table~\ref{tab_warm}). Lower panel: $N({\rm H})$ versus $kT_{in}$. Results obtained by \citet{kub07} and \citet{dia14} are included as well.}\label{logxi_nh_kt}
      \end{center}
   \end{figure} 
 
   \begin{figure} 
        \begin{center}
\includegraphics[scale=0.5]{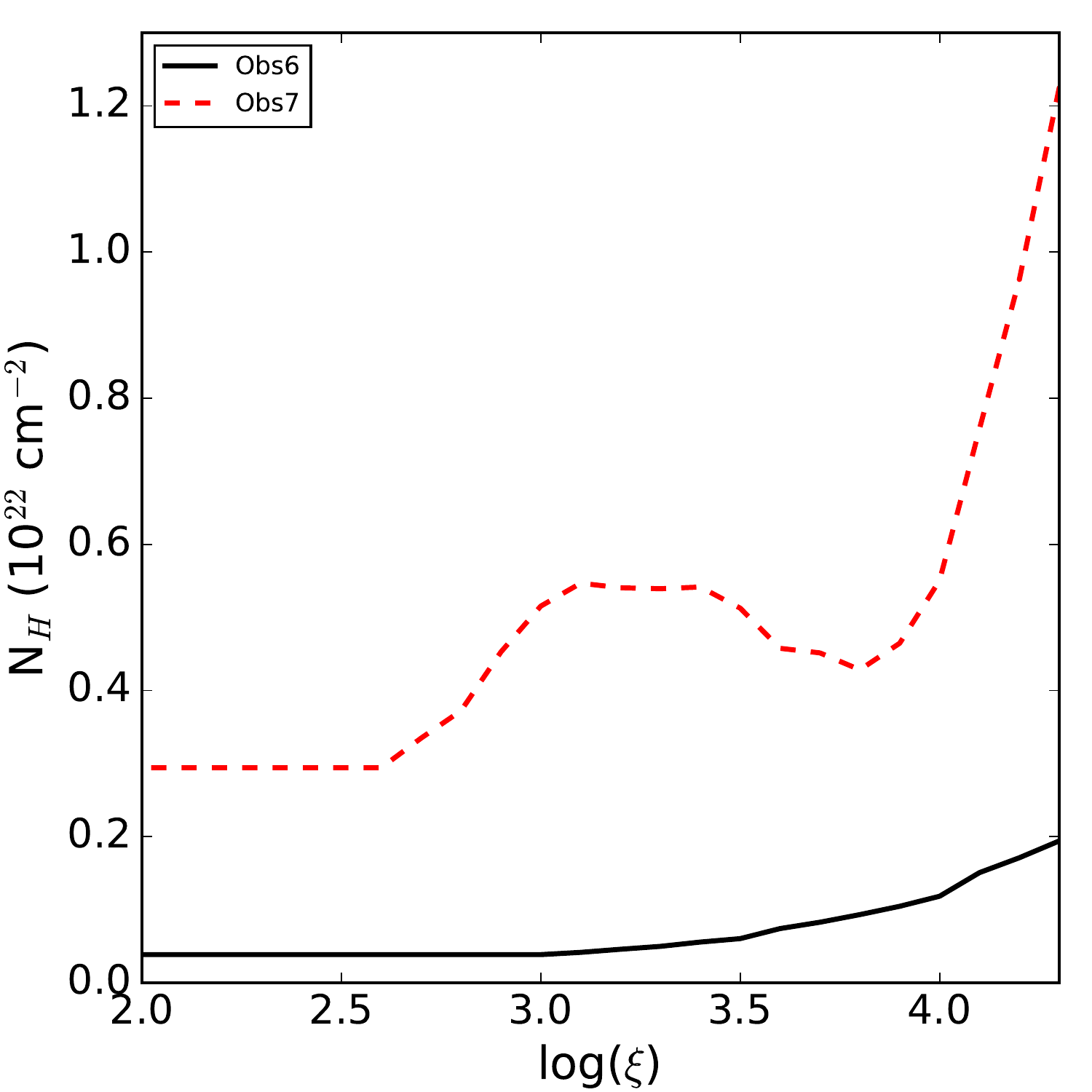}
      \caption{$N({\rm H})$ upper limits obtained with the {\tt warmabs} model for Obs~6 and 7 as function of $\log{\xi}$.}\label{limit_nh}
      \end{center}
   \end{figure} 
  
     \begin{figure*} 
        \begin{center}
\includegraphics[scale=0.32]{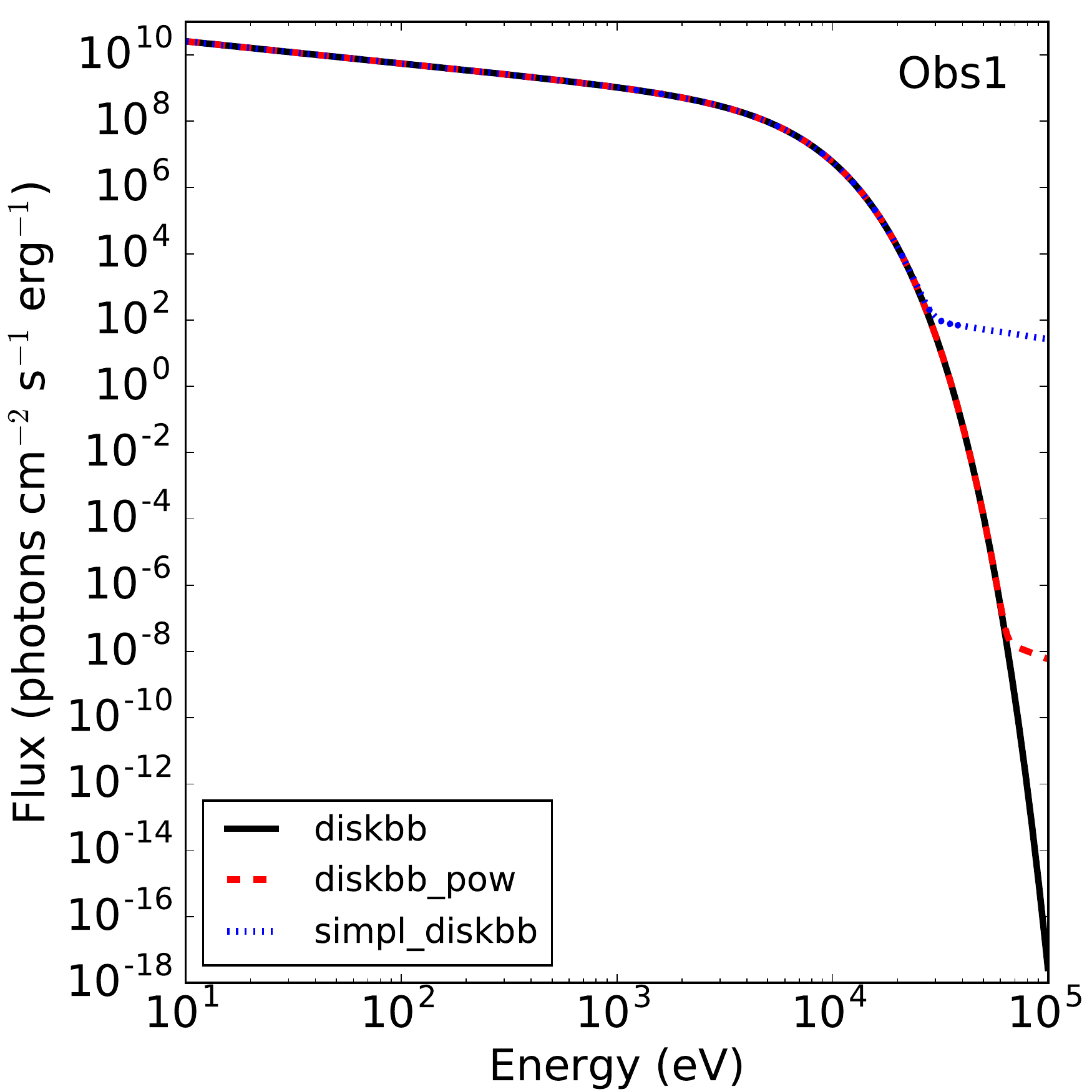}
\includegraphics[scale=0.32]{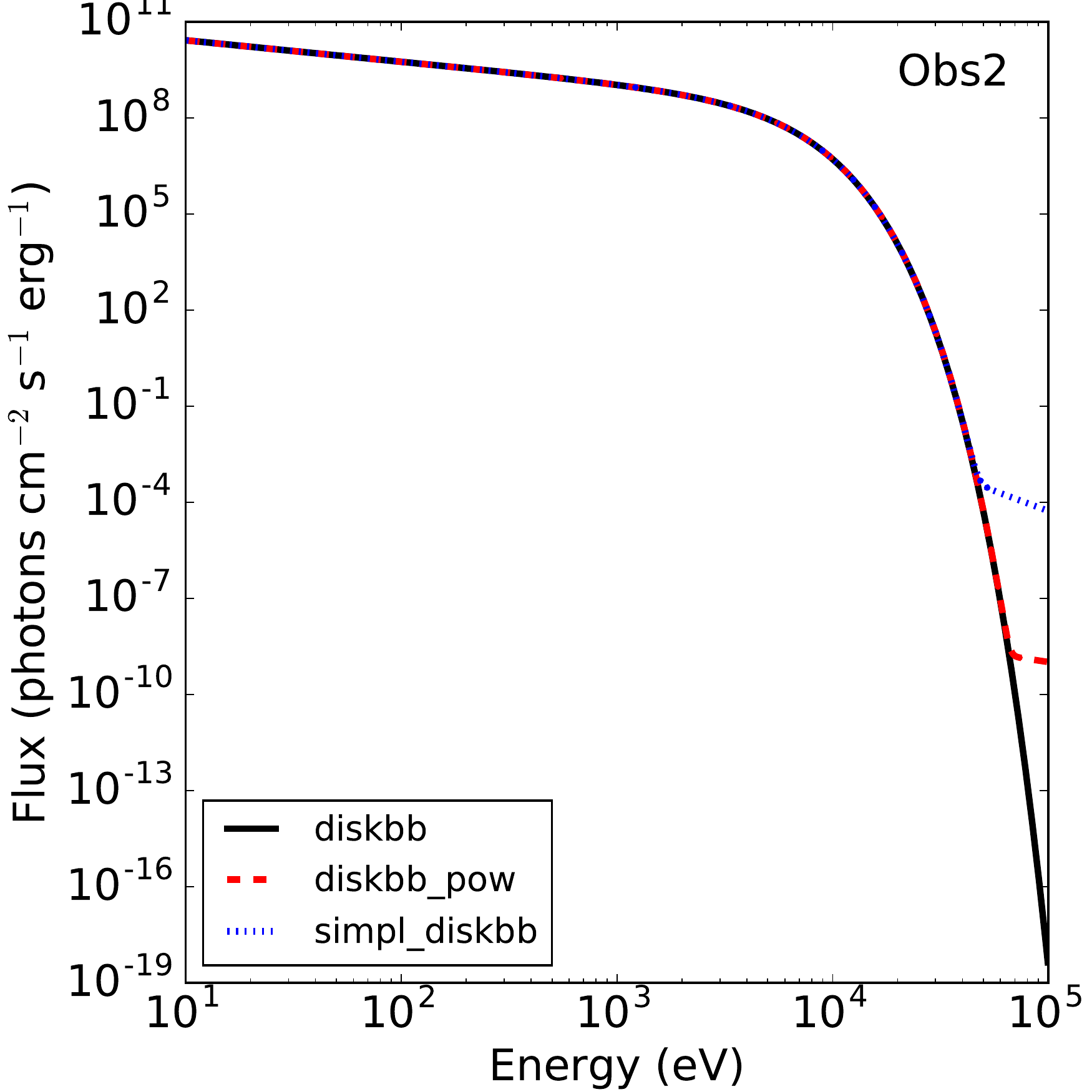}
\includegraphics[scale=0.32]{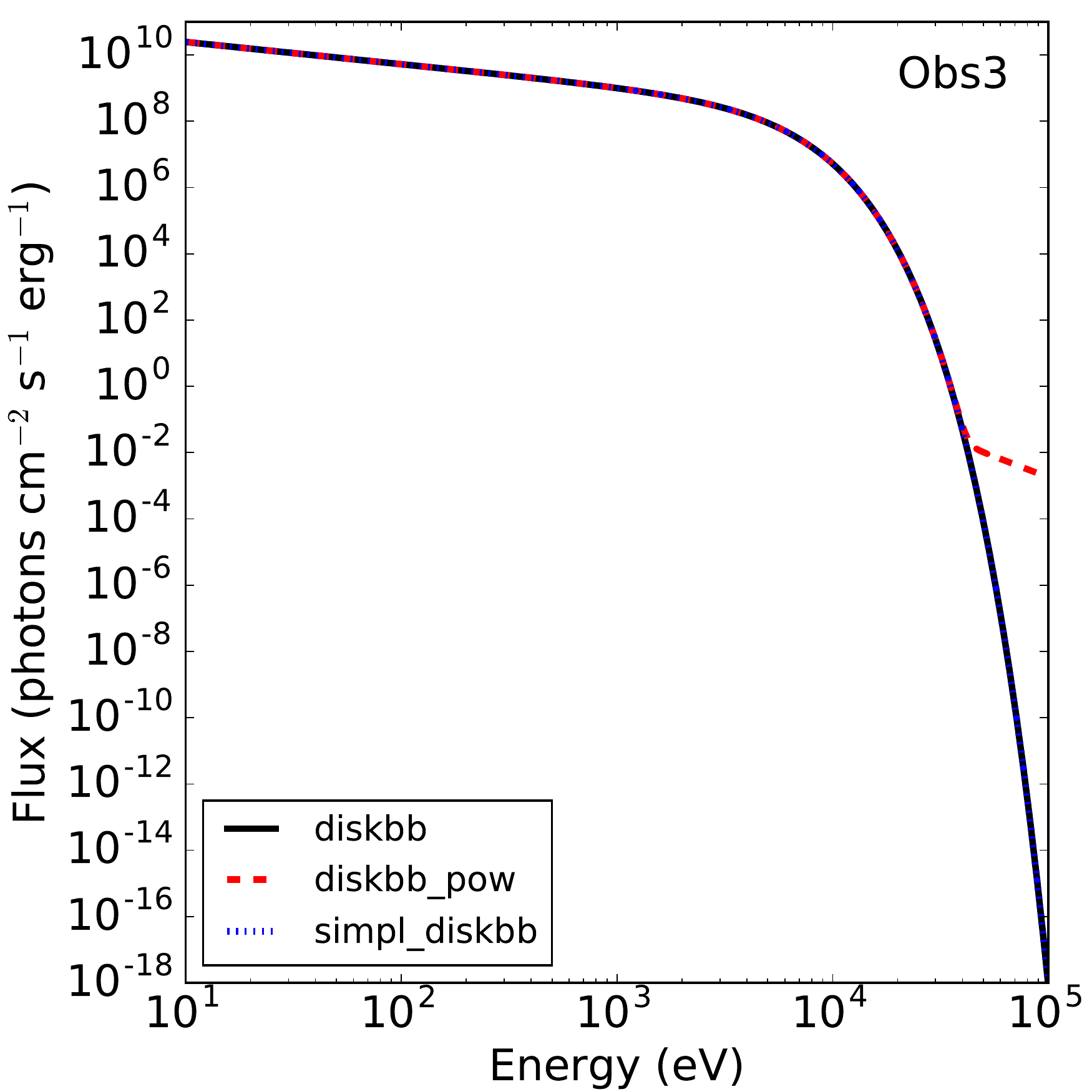}\\
\includegraphics[scale=0.32]{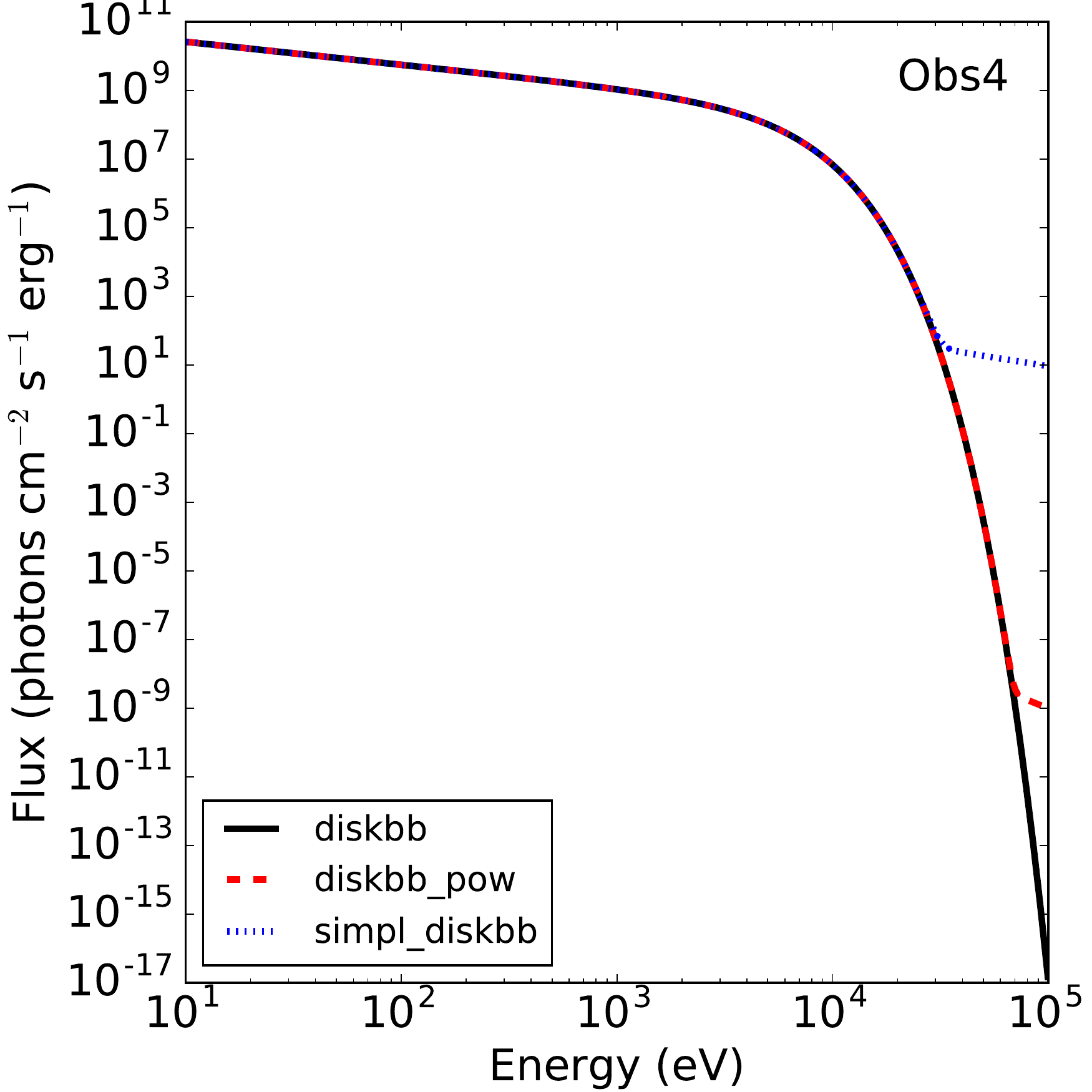}
\includegraphics[scale=0.32]{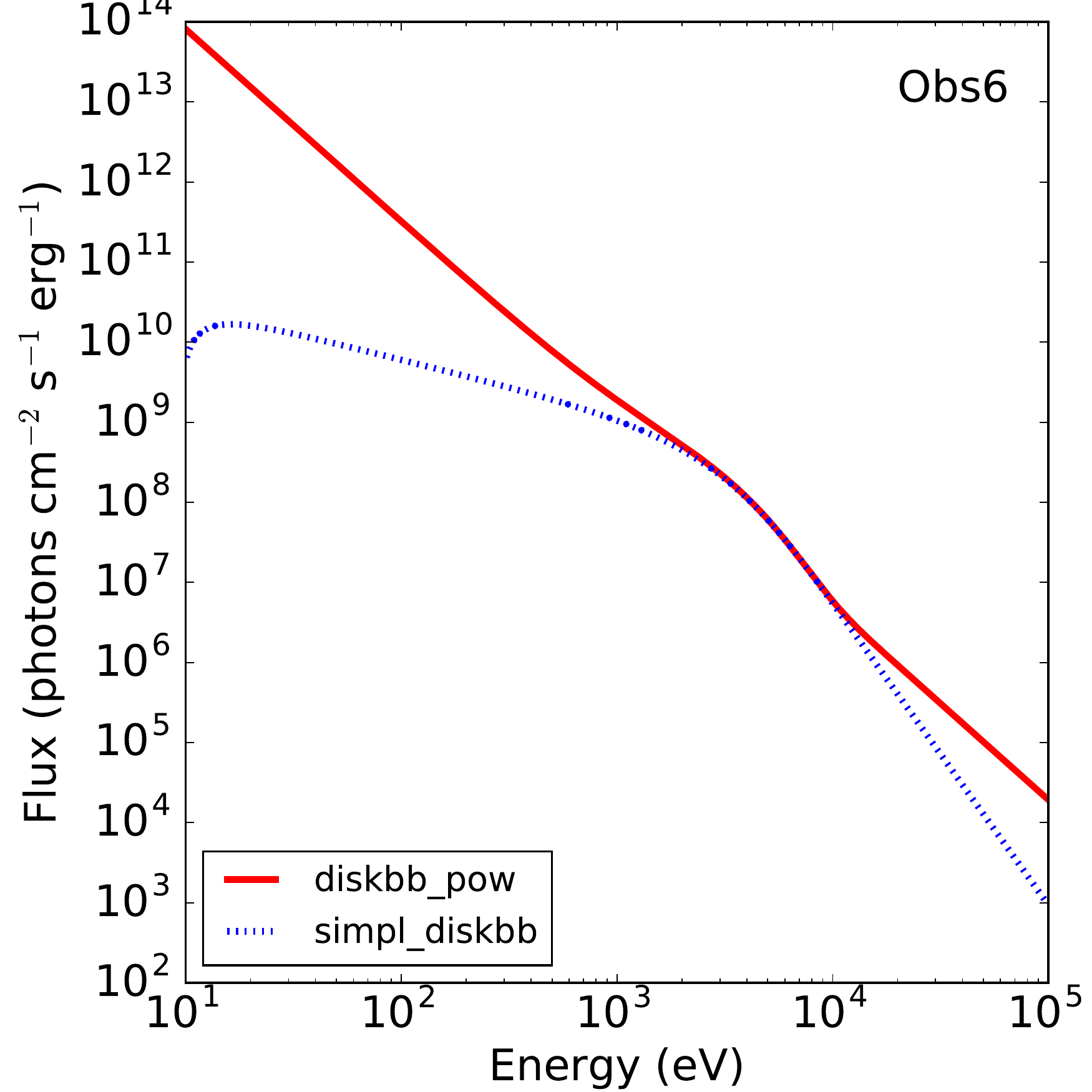} 
\includegraphics[scale=0.32]{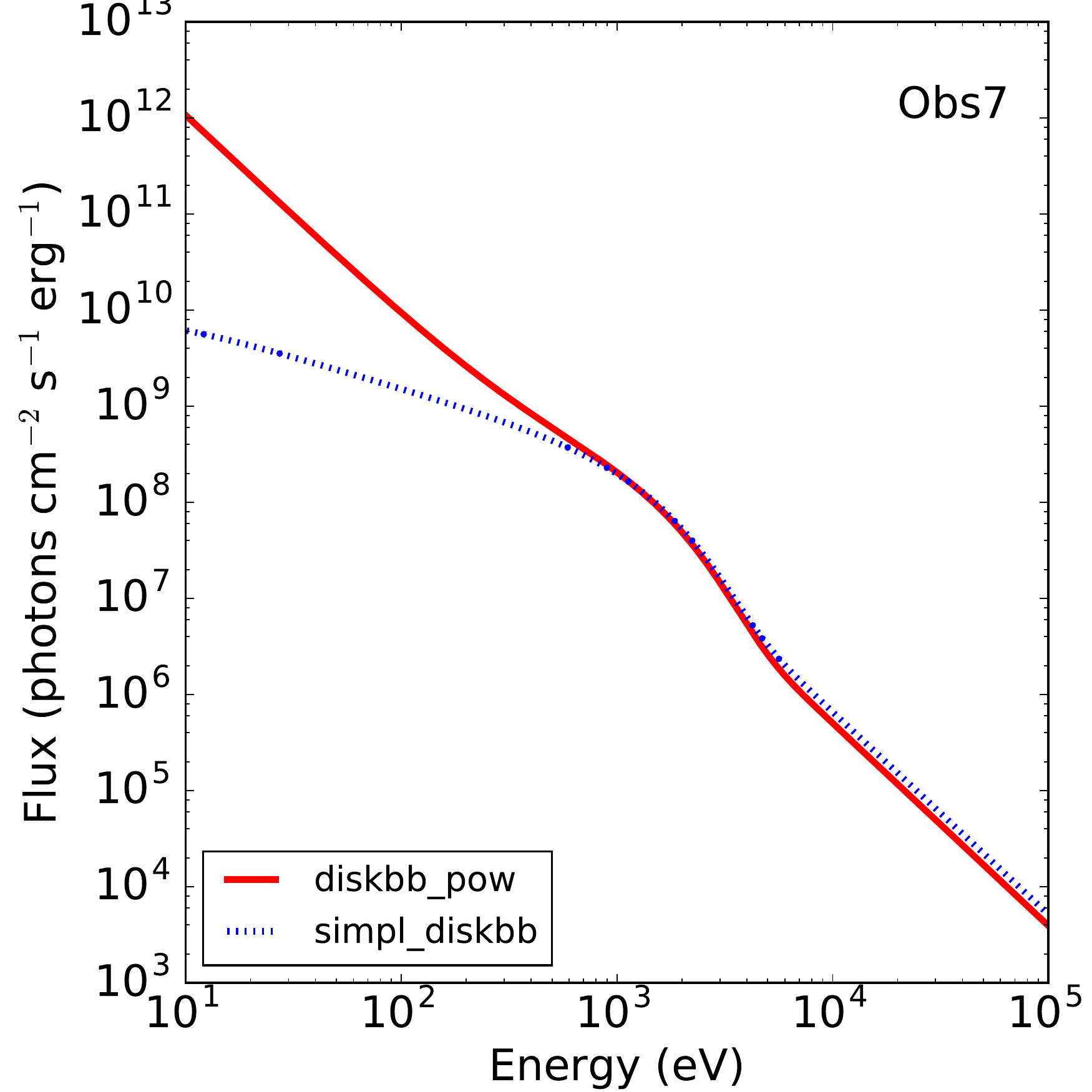}
      \caption{Spectral energy distributions obtained from the continuum modeling of Obs~1-4, 6 and 7.}\label{fig_seds1}
      \end{center}
   \end{figure*} 
  
   \begin{figure*} 
        \begin{center}
\includegraphics[scale=0.32]{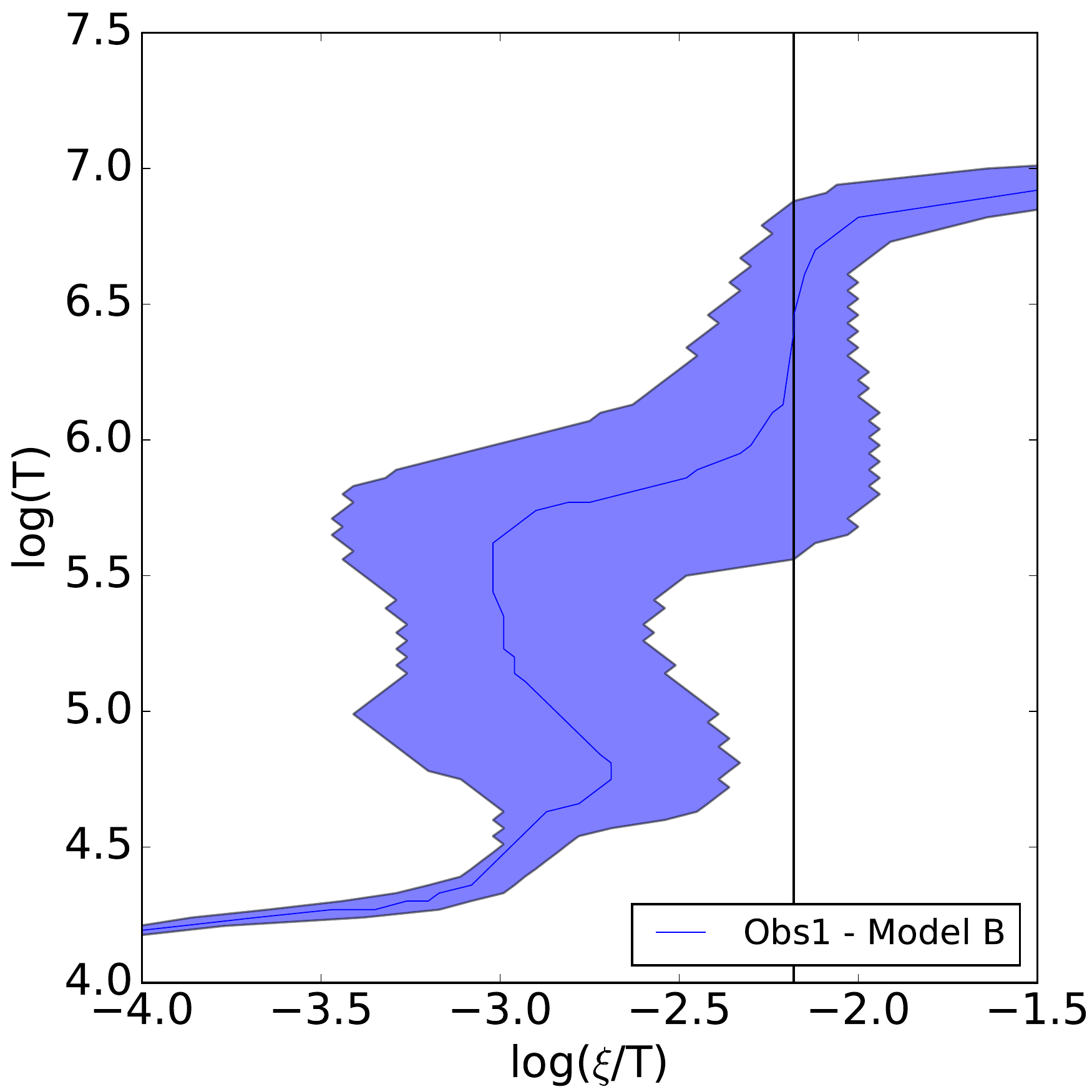}
\includegraphics[scale=0.32]{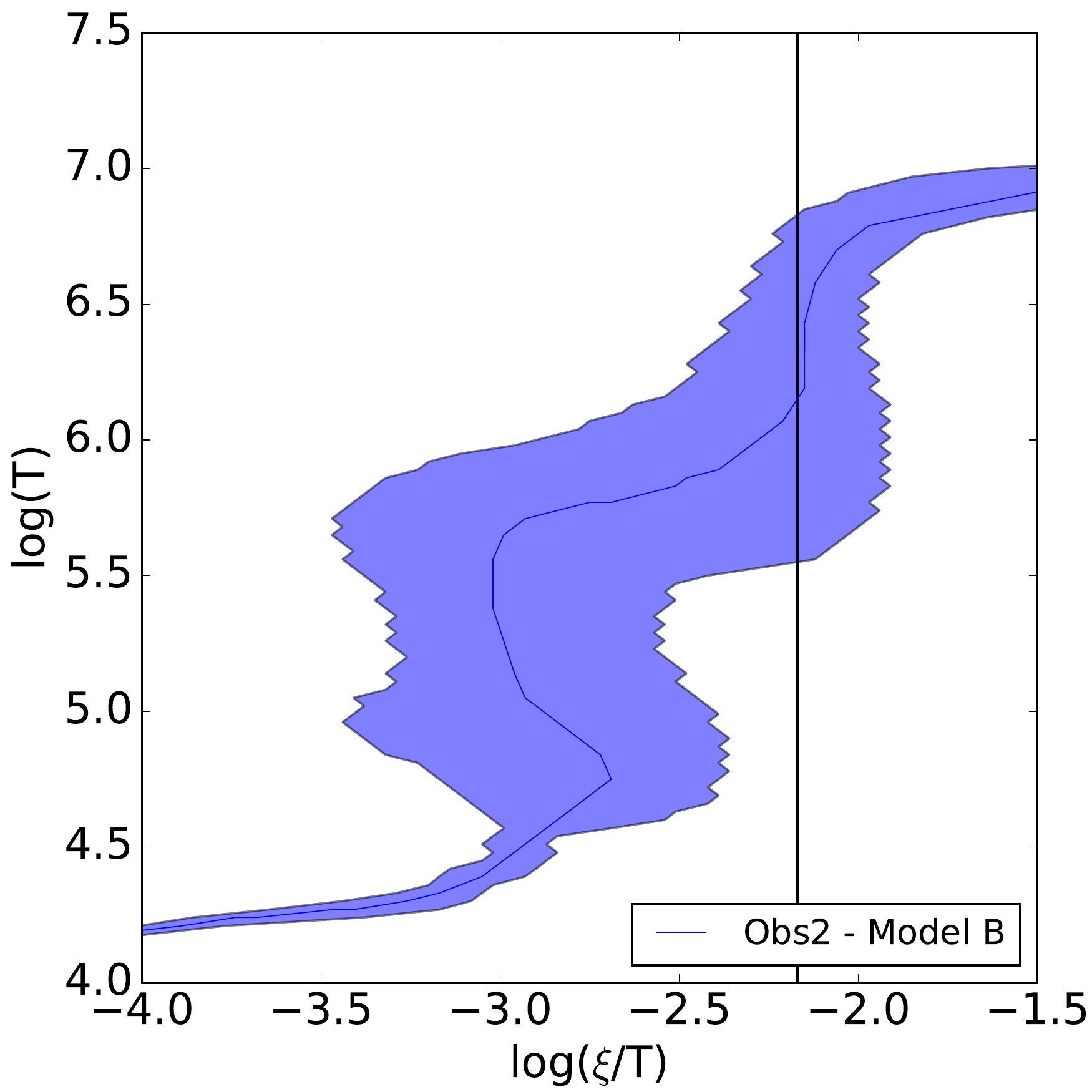}
\includegraphics[scale=0.32]{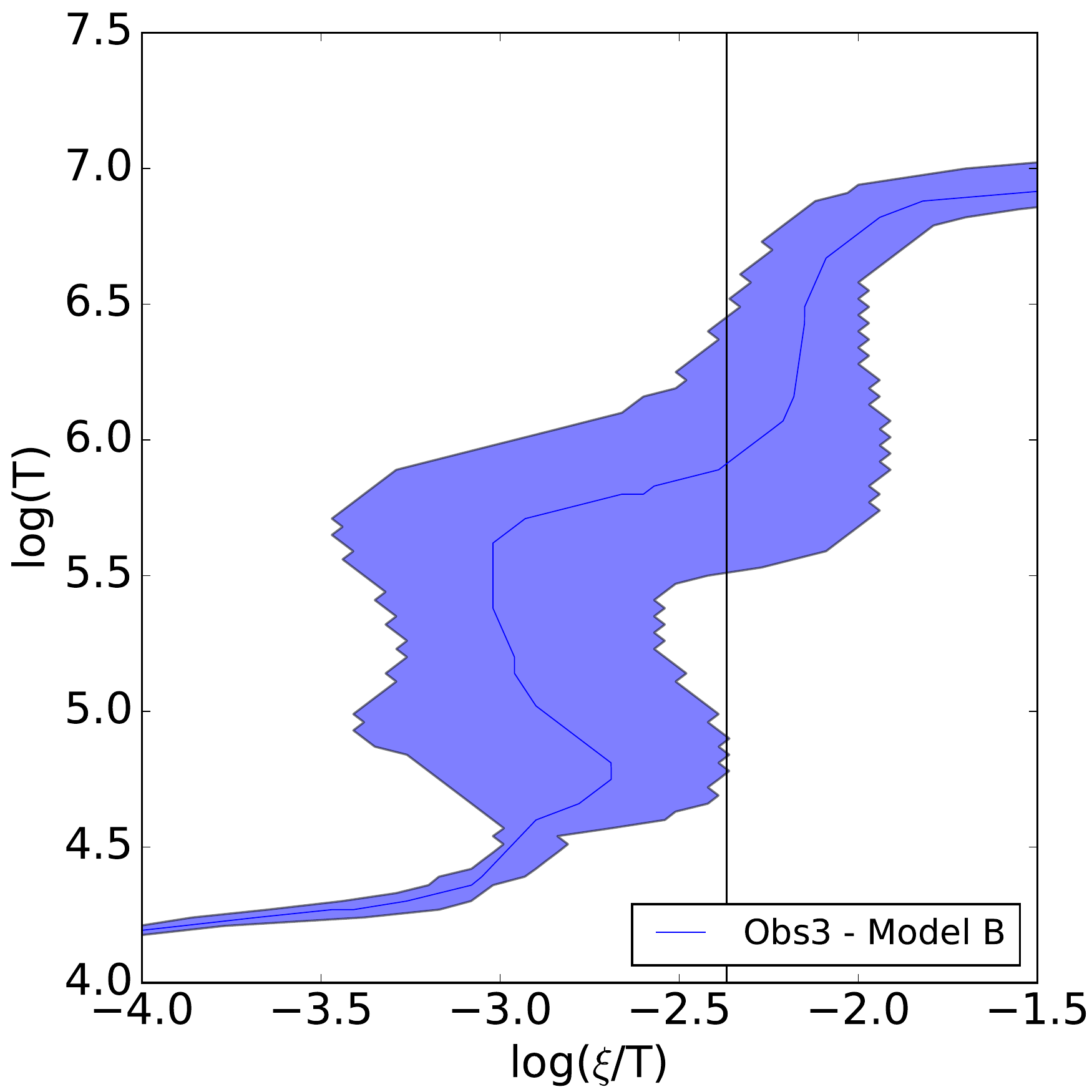}\\
\includegraphics[scale=0.32]{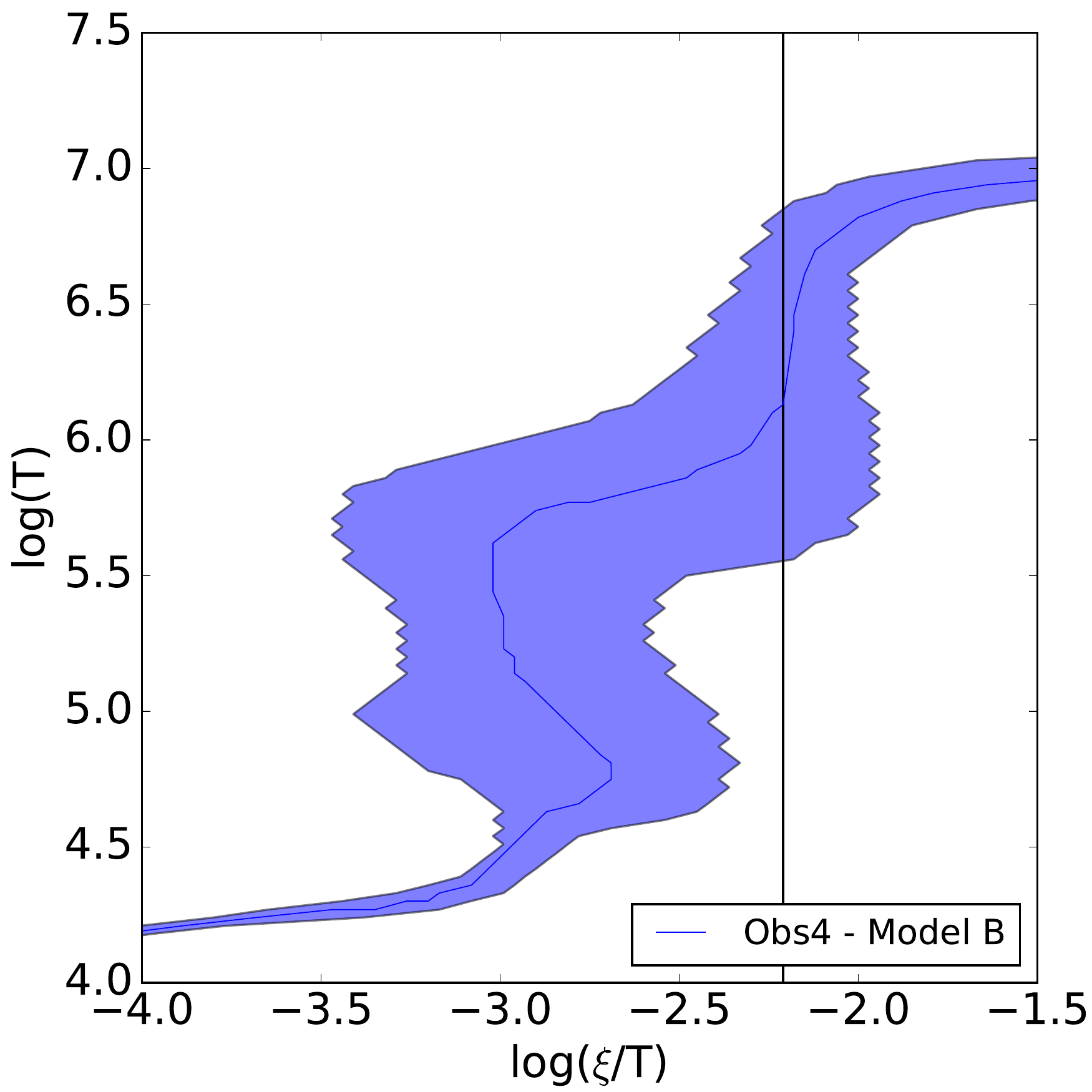}
\includegraphics[scale=0.32]{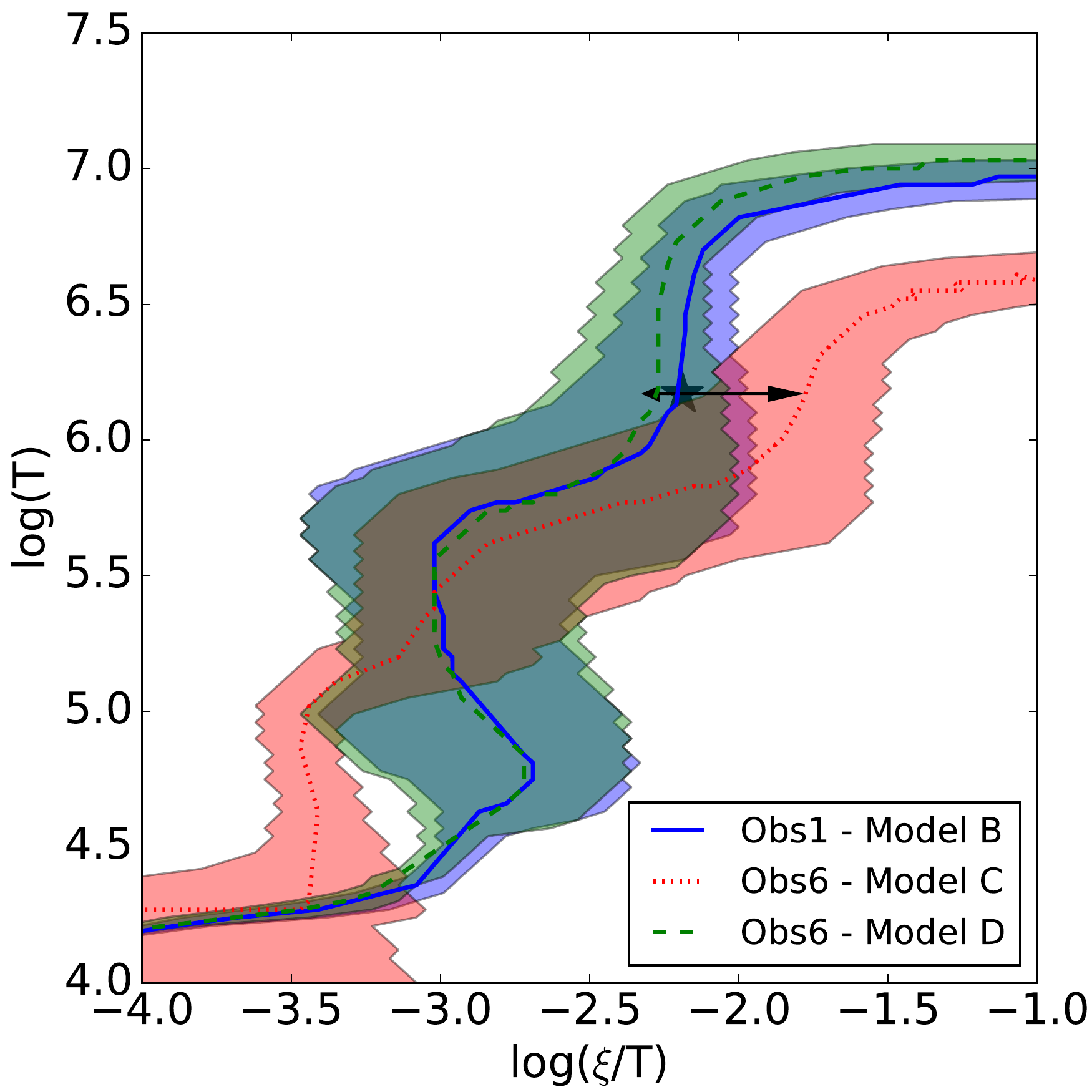}
\includegraphics[scale=0.32]{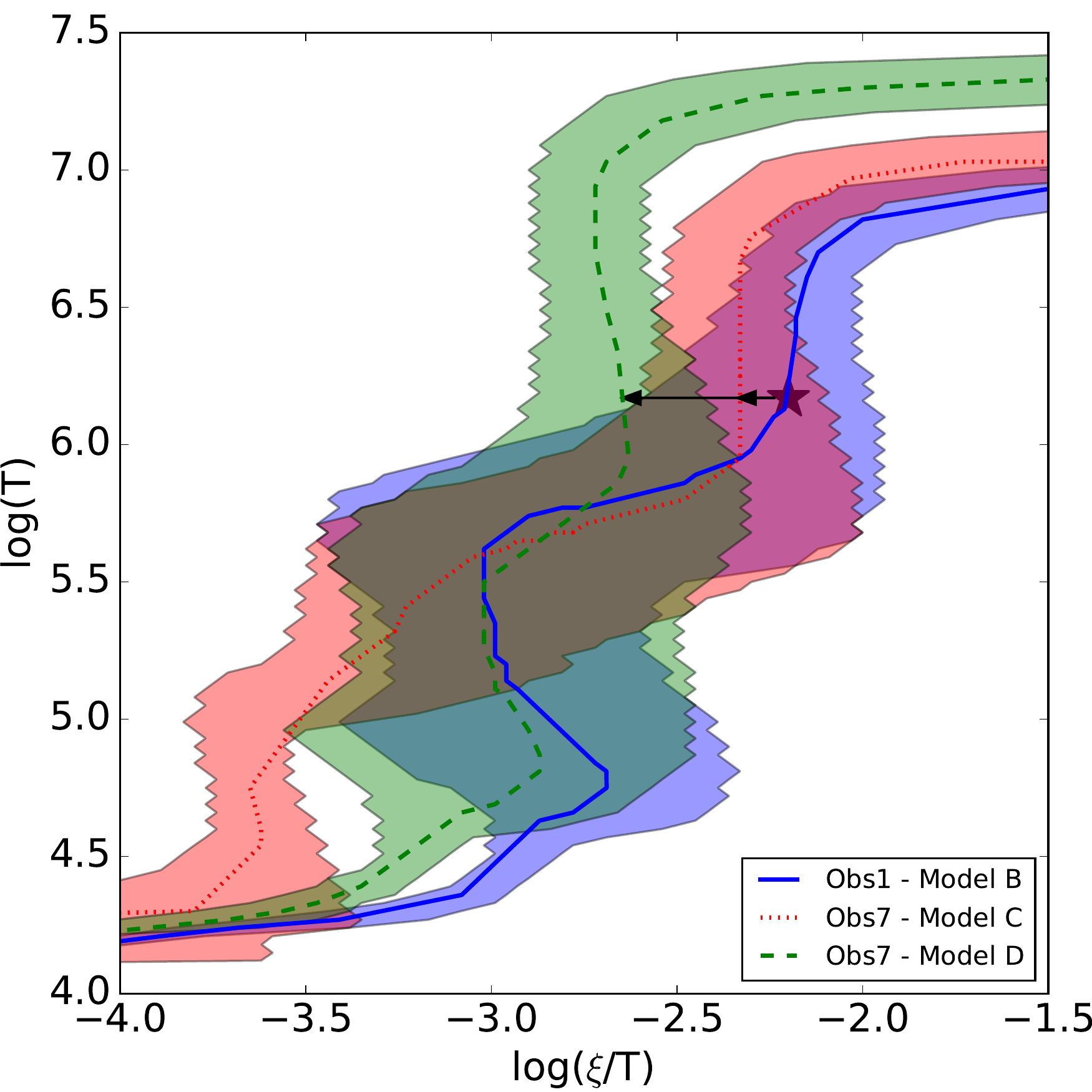}
      \caption{Thermal stability curves obtained for all observations. For Obs~1-4  the model B SEDs were used. Vertical lines correspond to the best fit parameters obtained with the {\tt warmabs} model. Middle and right lower pannels shows the results obtained for Obs~6 and 7 using Model C (red region) and Model D (green region). In both cases the black stars indicates the best fit parameters obtained for Obs~1 with model~B. Horizontal arrows show the possible transition between stability curves assuming $nr^{2}=$ constant. }\label{fig_ther_curve1}
      \end{center}
   \end{figure*} 
    
   \begin{figure} 
        \begin{center}
\includegraphics[scale=0.45]{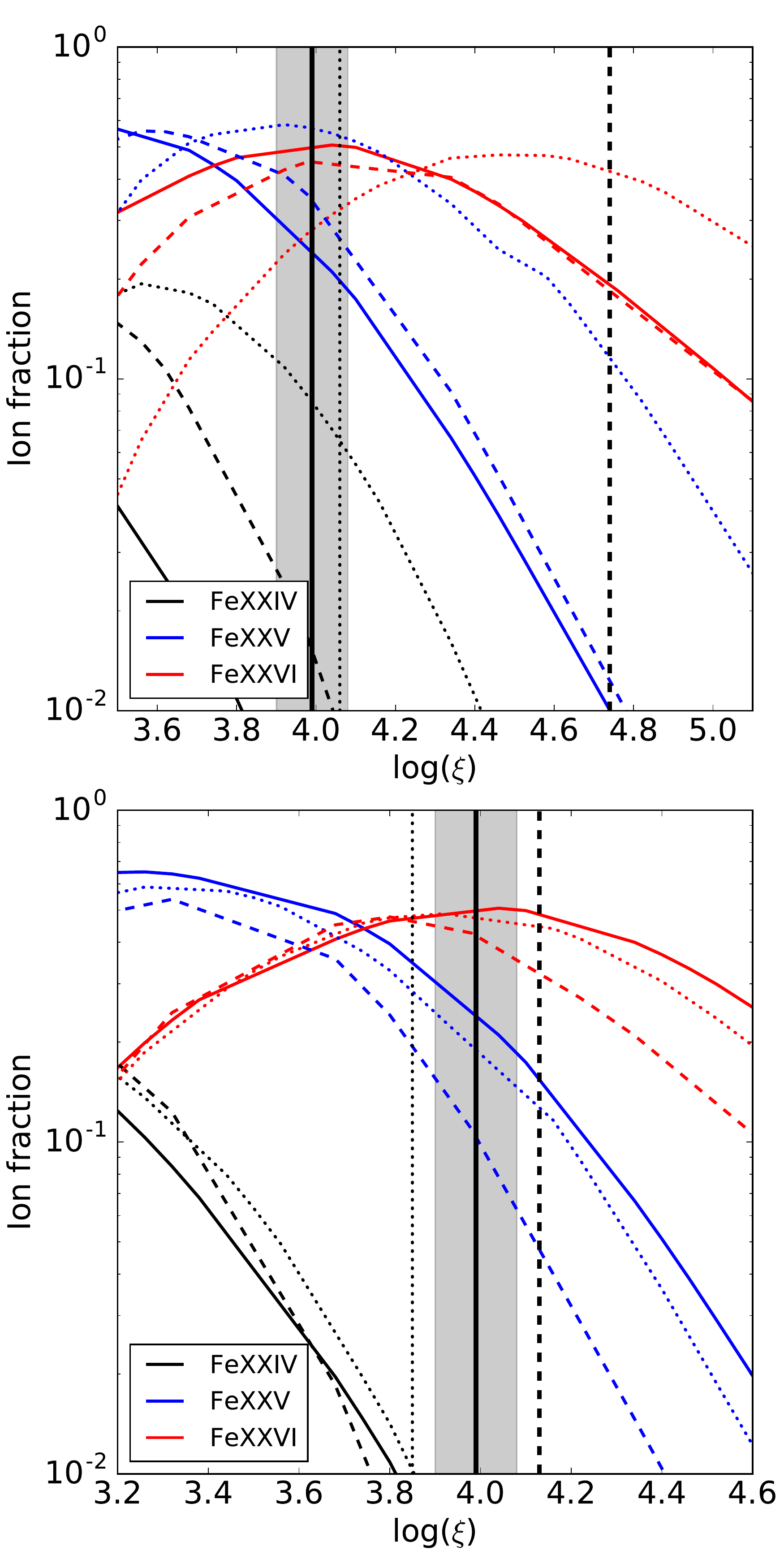}
      \caption{Ion fractions for the {\rm Fe}~{\sc XXIV}, {\rm Fe}~{\sc XXV} and {\rm Fe}~{\sc XXVI} ions as function of $\log{(\xi)}$ for Obs~1 (solid lines), Obs~6 (dotted lines) and Obs~7 (dashed lines). Top panel corresponds to the ion fractions obtained with model C for Obs~6 and 7 while bottom panel corresponds to the ion fractions obtained with model D for Obs~6 and 7. Vertical solid lines corresponds to the $\log{\xi}$  obtained for the best fit in Obs~1 including the uncertainty $\Delta\log{\xi}$ (gray region). The $\log{\xi}$ expected values obtained if the  $nr^{2}$=constant condition is assumed are also included for Obs~6 (dotted vertical line) and Obs~7 (dashed  vertical line), but see text for the caveats on this assumption.  }\label{fig_ion_fractions}
      \end{center}
   \end{figure}
 
\section{Observations and data reduction}\label{sec_dat}
\subsection{X-ray observations}

Table~\ref{tab_data} shows the specifications, including IDs, dates, exposure times and count rates, of the LMXB~4U~1630-47 {\it Chandra} observations analyzed in this paper. All observations were obtained using the Advanced CCD Imaging Spectrometer (ACIS) in combination with the High Energy Grating (HEG) of the High Energy Transmission Grating Spectrometer (HETGS) instrument. For each observation we combined both $\pm$ 1 HEG orders. The 1.5--10 keV {\it Chandra} count rates are included. The observations were reduced following the standard Chandra Interactive Analysis of Observations (CIAO, version 4.9) threads\footnote{\url{http://cxc.harvard.edu/ciao/threads/gspec.html}}.  The spectral fitting was performed with the {\sc xspec} software (version 12.9.1p\footnote{\url{https://heasarc.gsfc.nasa.gov/xanadu/xspec/}}), $\chi^{2}$ statistics were used and uncertainties are given at 90$\%$ confidence. Given that we are interested in the analysis of absorption features, we prefer to avoid the rebinning of the data due to the loss of information inherent to such a procedure. Instead, we prefer to use the \citet{chu96} weighting method which allows the analysis of low-counts spectra by assigning weights to each channel that correspond to the average counts in surrounding channels. In this way an almost unbiased accurate estimation of parameters is guaranteed but also a goodness-of-fit criterion is provided (contrary to, for example, the \citet{cas79} statistic). It is important to note that the smoothing of the data is performed only to calculate the weights, while the fitting procedure is applied to the original spectrum. This weighting method has been used in previous analyses of high-resolution X-ray spectra \citep{tof13,joa16,mil16c,gat18a,med18}. Finally, the abundances are given relative to \citet{gre98}.

\subsubsection{X-ray light curves}\label{sec_light}
Figure~\ref{fig_lc} shows the daily average lightcurves obtained during the 4U~1630-47 outburst with the Monitor of All-sky X-ray Image ({\it MAXI/ASM}) in the 2--20 keV and 10--20 keV energy ranges and the Neil Gehrels Swift Observatory with the Burst Alert Telescope ({\it Swift/BAT} in the 15--50 keV band (bottom panel).  In the case of {\it Swift/BAT} data, negative values, although without a physical meaning, are obtained from Poisson fluctuations of low significance bins due to the background subtraction method. Those values are not included in the plot.  Figure~\ref{fig_hr} shows the hardness-intensity diagram of the source using the {\it MAXI/ASM} daily average lightcurves, and where the hardness ratio is defined as the ratio between the observed fluxes in the 10-20 keV and the 2-4 keV bands. From the diagram, it is clear that Obs~6 is located in a similar region as Obs~1-4 while Obs~5 and 7 lie at a harder region of the diagram. In this sense, we use Obs~1-4 to compare with Obs~6-7. Given that Obs~5 does not show lines \citep{nei14} and it is yet in a particular state different from Obs~7, we decide to exclude it in the following analysis.
 
 \subsection{Radio observations}\label{sec_radio}
  During Obs~6 and 7 we made quasi-simultaneous radio observations with the Karl G.\ Jansky Very Large Array (VLA), under project code SE0242.  We observed in two basebands.  Each baseband comprised seven continuum spectral windows made up of sixty-four 2-MHz channels. We also centred an eighth spectral window in each baseband on the H94$\alpha$ and H110$\alpha$ recombination lines  \citep[rest frequencies of 7.792871\,GHz and 4.874157\,GHz, respectively, and assuming a systemic radial velocity of $-170$\,km\,s$^{-1}$, see][]{lil68}, to test for recombination line emission.  These spectral line frequencies were observed with narrower channels, of width 500\,kHz and 1\,MHz, respectively.

The array was in its most compact D configuration for Obs 6 (2013 April 25, 09:00-10:00 UT; MJD 56407.40$\pm$0.01), and the low declination of the source meant that the antennas on the northern arm were all shadowed and had to be flagged, leading to a very elongated synthesised beam.  We used 3C\,286 as a flux density and bandpass calibrator, and the more nearby calibrator J1626$-$2951 to set the complex gains.  We achieved a total of 35.3\,min of time on the target field.  However, the highly elongated synthesized beam and the large amount of diffuse structure in the field meant that the target source was not significantly detected, with a $3\sigma$ upper limit of 0.22\,mJy\,beam$^{-1}$ when stacking the two continuum basebands.

For Obs~7, the VLA observed on 2013 May 28 (06:46--07:46 UT; MJD 56440.31$\pm$0.01), using the same frequency setup and calibrator sources.  However, the array was in the slightly more extended DnC configuration, providing slightly better N-S resolution.  Once again, the diffuse emission in the field coupled with the low elevation of the source, the compact array configuration and the relatively poor N-S resolution hampered our ability to detect the target.  We experimented with both different weighting schemes and minimum {\it uv}-distance restrictions on the data, but we were only able to place $3\sigma$ upper limits of 0.48\,mJy\,beam$^{-1}$ at 5.3\,GHz and 0.12\,mJy\,beam$^{-1}$ at 7.2\,GHz (where the diffuse emission was fainter and we spatially resolved out a fraction of the more extended structure). In neither observation did we detect any significant recombination line emission (rms noise levels per channel of 0.47 and 0.69\,mJy\,beam$^{-1}$ for H94$\alpha$ and H110$\alpha$, respectively).

\section{X-ray spectral modeling}
\subsection{Continuum modeling}\label{sec_cont}
In order to account for different spectral states we fitted each HEG-ACIS observation in the 1.5--10 keV energy range using multiple phenomenological models. Using {\sc xspec} nomenclature, the models are:
\begin{enumerate}
\item Model A: {\tt tbabs*(powerlaw)}
\item Model B: {\tt tbabs*(diskbb)}
\item Model C: {\tt tbabs*(powerlaw+diskbb)}
\item Model D: {\tt tbabs*simpl(diskbb)}
\end{enumerate}
  
The {\tt diskbb} component corresponds to an accretion disc consisting of multiple blackbody components \citep{mit84,mak86}. The {\tt simpl} convolution component is a model of comptonization in which a fraction of the photons (FracSca parameter) is scattered into a power-law component \citep{ste09}. The {\tt tbabs} component models the absorption in the local interstellar medium (ISM) as described by \citet{wil00}. In this way, models A and B consider X-ray spectra dominated by hard and soft components, respectively,  while models C and D correspond to a hybrid case.  Also, we included several {\tt gaussians} in order to model absorption lines identified in the spectra, if present. 

The pileup effect, the detection of two or more photons as a single event, can affect the shape and level of the continuum\footnote{\url{http://cxc.harvard.edu/ciao/ahelp/acis\_pileup.html}}. The pileup effect is stronger in the Medium Energy Grating (MEG) than in the HEG instrument. Currently, the only model that can be used to estimate the pileup effect for high-resolution X-ray spectra is  {\tt simple\_gpile2} developed by \citet{han09} for the {\sc isis} data analysis package. Using this model, we estimated the highest pileup degree in Obs~1-4, 6-7 to be $<5\%$ at $\approx 3.8$ keV. We have found that the continuum parameters listed are not affected by the inclusion of the {\tt simple\_gpile2} model and hence we decide to use the results obtained from the {\sc xspec} analysis.  
   
Table~\ref{tab_con} shows the best-fit parameters obtained for each observation using the models described above. All Gaussians included in the models are listed in Table~\ref{tab_gauss}. We consider as valid fits those satisfying (1) $\chi^{2}$/dof $<$ 1.50 and (2) the number of counts predicted by the model in the 15-50 keV energy range agrees with the {\it Swift/BAT} daily average measurements (within the uncertainties). Models that do not satisfy both conditions are not included in the analysis hereafter (including the Model~A).

It is important to note that, because we do not have hard-energy range spectra ($>$ 10 keV), the {\tt powerlaw} photon-index cannot be well constrained because the fit can estimate a large, unrealistic, photon-index by increasing the absorption at low energies.  Consequently, when using model C to fit Obs~1-4 we found that the $\Gamma$ parameter artificially increases to values $>$ 5 and  therefore we decide to fix it to an acceptable value of $2.5$. \citet{sei14} reported $\Gamma$ values up to 3 by analyzing {\it RXTE} observations along multiple accretion states, although given that our $\Gamma$ value obtained from the more physical model {\tt simpl} is $<2.0$ we have considered that $\Gamma=3.0$ is a bit too large. On the other hand, the fit obtained using models C, D for Obs~7 tend to decrease $N({\rm H})$ to unrealistic values (i.e $< 0.01$ $\times 10^{22}$ cm$^{-2}$). In these cases, we decide to fix $N({\rm H})$ to the minimum value obtained from the best-fits in Obs~1 (i.e. $9.10$ $\times 10^{22}$ cm$^{-2}$ obtained with model B and considering the uncertainty).
 
Figure~\ref{fig_dat_1} shows the best-fit models and residuals for all observations analyzed. Data were rebinned for illustrative purposes.  For Obs~1-4 Gaussians were included to account for multiple absorption lines visible in the spectra (see Section~\ref{sec_lines}).  Absorption lines were not detected in Obs~6-7.  In all figures, lower panels indicate the data/model ratios obtained for the different models described in Table~\ref{tab_con}.
  
Given the results obtained from the continuum modeling, in combination with the {\it Swift/BAT} fluxes, we classify the observations into accretion states as follows. Obs~1-4 are best modeled with a {\tt diskbb} with a relatively high temperature (1.49--1.58 keV) and show no significant hard X-ray flux (15--50 keV). Based on this, on their intensities and spectral hardnesses, and on their similarity to the {\it XMM-Newton} observations analyzed in \citet{dia14}, we classify them as being in a soft accretion state. 

In contrast, Obs~6 and 7 require two-component models ({\tt diskbb} and a {\tt powerlaw}) to predict the {\it Swift/BAT} flux within the errors. The temperature of the disc decreases steadily from Obs~1-4 ($kT$ $\sim$ $1.5-1.6$ keV) to Obs~6 ($kT$ $\sim$ $0.9-1.2$ keV) and Obs~7 ($kT$ $\sim$ $0.6$ keV). In contrast, the normalization of the power law representing the hard X-ray emission first increases from Obs~1-4 to 6 by a factor $>30$, highlighting the departure of Obs 6 from a soft state. Obs~7 shows a drop in total luminosity of one order of magnitude with respect to Obs~6. The decrease of luminosity is accompanied again by an increase of the power law fraction and therefore the hardness ratio, as indicated in Figure~\ref{fig_hr}. We have estimated the power law contribution to the total unabsorbed flux in the 2--20 keV energy range for Obs~7 to be $\sim 65\%$  and $\sim 62\%$ for Models C and D, respectively. In the case of Obs~6 we have estimated the powerlaw contribution to the total unabsorbed flux in the 2--20 keV energy range to be $\sim 51\%$  and $\sim 49\%$ for Models C and D, respectively. As a reference, \citet{mcc06} indicates that the powerlaw contribution should be $<25\%$ and $>80\%$ in order to be classified as a soft and hard accretion state, respectively. Following this prescription, both Obs~6 and~7 would correspond to intermediate states. However, we emphasize that the restricted energy band imposes some limitations on our fits. As an example, we show in Figure~\ref{fig_contour} a contour map of the $N({\rm H})$ and the $\Gamma$ parameters obtained from Model~C for Obs~6 and Obs~7. From this plot it is clear that Obs 7 favors harder photon index compared to Obs~6. Taking all the above into account we classify Obs~6 as being a relatively soft-intermediate state and Obs~7 as a likely hard-intermediate or hard state. Assuming a black-hole mass of 10 \(\textup{M}_\odot\) we have found that all observations analyzed correspond to to sub-Eddington luminosities.

\subsection{Absorption lines}\label{sec_lines}

Table~\ref{tab_gauss} shows the absorption lines identified in Obs~1-4, including their energies (keV), widths ($\sigma$), fluxes (photons cm$^{-2}$ s$^{-1}$) and equivalent widths (EWs).  Common lines for Obs~1-4 include {\rm Fe}~{\sc XXVI} K$\alpha$,K$\beta$, {\rm Fe}~{\sc XXV} K$\alpha$,K$\beta$, {\rm Ca}~{\sc XX} K$\alpha$. On the other hand, {\rm Ar}~{\sc XVIII} K$\alpha$, {\rm S}~{\sc XVI} K$\alpha$, {\rm Si}~{\sc XIV} K$\alpha$ absorption lines are identified only in Obs~3 (see Table~\ref{tab_con}). We have not detected significant absorption lines in Obs~6-7, and we estimated upper limits of 1 eV for {\rm Fe}~{\sc XXVI} K$\alpha$ and {\rm Fe}~{\sc XXV} K$\alpha$, respectively (see Table~\ref{tab_gauss}). 
 
Considering the uncertainties of the measurements we cannot identify significant changes in the line positions and EWs among the observations, except for {\rm Fe}~{\sc XXV} K$\alpha$ in Obs~3. We found that the blueshift of the lines ranges from 200 to 500~km/s between the different observations, consistent with values reported by \citet{nei14}. We noted that the change in the energy position as well as the increase of the EW for the {\rm Fe}~{\sc XXV} K$\alpha$ line in Obs~3 is probably due to a blending of the line with {\rm Fe}~{\sc XXIV} K$\alpha$ located at $\sim 6.690$ keV.   The {\rm Fe}~{\sc XXV} K$\alpha$/K$\beta$ and {\rm Fe}~{\sc XXVI} K$\alpha$/K$\beta$ line ratios correspond to a plasma with $N({\rm H})\gtrsim 10^{23}$ cm$^{-2}$ and $\nu_{turb}\gtrsim 200$ km/s \citep[table 3 of][]{roz06}.  Finally, the EWs distribution as function of the flux for the {\rm Fe}~{\sc XXV} and {\rm Fe}~{\sc XXVI} ions suggest that both K$\alpha$ lines are saturated, in which case their width (i.e. their $\sigma$) may be over-estimated. However the uncertainties of the line property measurements do not allow to be conclusive in this respect.

\subsection{Photoionization modeling}\label{sec_dis}
Having obtained a phenomenological description of the spectra, we substitute the Gaussian components in Obs~1-4 described in Section~\ref{sec_cont} with the photoionization model {\tt warmabs}\footnote{https://heasarc.gsfc.nasa.gov/xstar/docs/html/node102.html}. This model is part of the {\sc xstar}\footnote{http://heasarc.nasa.gov/lheasoft/xstar/xstar.html} photoionization code which is designed to compute the physical conditions for an ionizing source surrounded by a gas taking into account physical processes such as photoionization, electron impact collisional ionization and excitation, and radiative and dielectronic recombination.  The main assumptions include ionization equilibrium conditions, a Maxwellian electron velocity distribution and that the gas responsible for absorption and emission has an uniform ionization and temperature throughout. The {\tt warmabs} model parameters include the column density of the absorber ($N({\rm H})$), ionization parameter ($\log{\xi}$), elemental abundances (A$_{x}$), broadening turbulence ($v_{turb}$), and redshift ($z$). In order to perform a self-consistent modeling, we used the unabsorbed spectral energy distributions (SEDs) obtained from the continuum fits in Section~\ref{sec_cont} as the central source of ionizing radiation to compute the energy level populations required by {\tt warmabs}.  
 
Table~\ref{tab_warm} shows the best-fit parameters obtained for Obs~1-4 using the same continuum as for model~B (i.e. {\tt tbabs*warmabs*diskbb}). Abundances were fixed to solar values for both the ISM component and the ionized absorber. As the $\chi^2_{\rm red}$ values indicate, this model slightly improves the fit as compared to the Gaussians described in Section~\ref{tab_con}. Figure~\ref{iron_warm} shows the best fit obtained with the {\tt warmabs} model. The main absorption features/lines included in the model are indicated. The column densities derived for {\rm Fe}~{\sc XXV} and {\rm Fe}~{\sc XXVI} with {\tt warmabs} are $N({\rm H})> 10^{23}$ cm$^{-2}$, in agreement with the line ratios obtained from the fit with Gaussians. For Obs~3, residuals around {\rm Fe}~{\sc XXVI} K$\alpha$ and {\rm Fe}~{\sc XXVI} K$\beta$ may indicate the presence of a second plasma component or line saturation. However, we did not find a better fit (i.e. an improvement in the statistic) by adding a second {\tt warmabs} component. The best-fit {\tt warmabs} parameters  are very similar between the observations except for the lower $\log{(\xi)}$ and a lower blueshift in Obs~3. In this sense, \citet{hig15} have shown that variations in the luminosity and/or plasma density affect the maximum blueshift of the disc wind. 

Figure~\ref{ng_logxi} shows a comparison between the $\log{(\xi)}$, $N({\rm H})$ and fluxes obtained from the best-fits. It is clear that Obs~3, which shows lower flux in the 0.013--13.6 keV energy range, requires  both a lower column density and lower ionization parameter to be modeled. The decrease of $\log{(\xi)}$ to $\sim 3.55$ leads to the appearance of absorption lines associated to ions at lower ionization state, such as {\rm Ar}~{\sc XVIII} and {\rm S}~{\sc XVI}  (see Figure~\ref{fig_dat_1}).

The wind launching radius can be estimated from the ionization parameter obtained with {\tt warmabs} for Obs~1-4 through the well known relation $\xi = L/nr^{2}$ where $L$ is the total luminosity in the 0.013--13.6 keV energy range, $n$ is the hydrogen plasma density and $r$ is the radius of the innermost edge of the shell surrounding the source \citep{tar69}. Assuming $n=10^{12}$ cm$^{-3}$, we obtain values for the radius of ($1.3\pm 0.1$)($10^{12}$ cm$/n$)$ \times 10^{11}$~cm  (Obs~1), ($1.3\pm 0.2$)($10^{12}$ cm$/n$)$\times  10^{11}$~cm  (Obs~2), ($2.1\pm 0.5$)($10^{12}$ cm$/n$)$\times 10^{11}$~cm  (Obs~3) and ($1.4\pm 0.2$)($10^{12}$ cm$/n$)$\times  10^{11}$~cm  (Obs~4). The $n=10^{12}$ cm assumption is derived from the measuring the luminosity of the source and column density and ionization of the absorber and making the assumption that $\Delta R/R \sim 1$, i.e. that the thickness of the absorber is similar to its radius \citep[see][sect. 4.3 for such a derivation]{kub07}.   Using the same $n$, \citet{kub07} and \citet{dia14} estimated a launching radius down to one order of magnitude lower. In this sense, discrepancies are due to differences in the total luminosity obtained from the SEDs and the $\xi$ values.

Figure~\ref{logxi_nh_kt} shows $\log{\xi}$ and $N({\rm H})$ versus $kT_{in}$ obtained for Obs~1-4. For comparison, results obtained by \citet{kub07} and \citet{dia14} are included as well, which correspond to {\it Suzaku} and {\it XMM-Newton} observations, respectively. The plot shows that even for similar $kT_{in}$, differences can be found in the $\log{\xi}$ and $N({\rm H})$ parameters obtained. Figure~\ref{limit_nh} shows $N({\rm H})$ upper limits obtained with the {\tt warmabs} model for Obs~6 and 7 as function of $\log{\xi}$. In both cases the continuum corresponds to model C. Clearly, the column density upper limits are very restrictive for Obs~6, with a maximum column density of  $2\times 10^{21}$~cm$^{-2}$ for a $log(\xi)$ of 4.2 and even lower column densities of  $5\times 10^{20}$~cm$^{-2}$ for a $log(\xi)$ < 3. For Obs~7 the column density upper limits are less constraining, especially for $log(\xi)$ $\gtrsim$ 3.8 for which column densities as high as  $1.2\times 10^{22}$~cm$^{-2}$ could be present. For $log(\xi) < 3.8$, the column density of a potential absorber is $<5\times 10^{21}$~cm$^{-2}$.     
 
\section{Stability curves}\label{sec_ther}
The equilibrium states of a photoionized plasma can be studied through the stability curve (or thermal equilibrium curve), which consists of a $T$ versus $\xi/T$ diagram \citep{kro81}. When the plasma reaches a state outside the stability curve the heating and cooling processes will compete until reaching the equilibrium. Depending on the slope of the stability curve, we can identify parts of the curve during which the slope is positive, corresponding to thermally stable regions, and negative, corresponding to thermally unstable regions. 

We created stability curves using the {\sc xstar} photoionization code (version 2.41) to analyze the equilibrium conditions for the plasma associated to 4U~1630-47. We ran a grid in the ($log(T),log(\xi)$) parameter space, with values ranging from $4 < \log{(T)} <10$  and $-4 < \log{(\xi)} <8$. We assumed an optically thin plasma with a constant density $n=10^{12}$ cm$^{-3}$ and solar abundances. For each ($log(T),log(\xi)$) point, the heating and cooling rates, as well as the ionic fractions for all elements, are stored. In this way, we can determine the ($log(T),log(\xi)$) values corresponding to a thermal equilibrium state (i.e. heating = cooling). 
 
Because the stability curves are strongly affected by the shape of the SED \citep[see for example][]{kro81,cha09} we study the effects of all continuum models described in Section~\ref{sec_cont} as ionizing continuum in the {\sc xstar} grid calculation.  Figure~\ref{fig_seds1} shows the SEDs for the different observations and models used to generate the stability curves. For Obs~1-4, different models yield differences in the SEDs only in the high energy region (consistent with the continuum degeneracy found in the {\it Chandra} spectra, see Table~\ref{tab_con}). In a way, we are forcing the difference in Obs~7 to be only at low energies by fixing the $N({\rm H})$. As explained in Section~\ref{sec_cont}, we found that for Obs~6 the model can yield dramatically different SEDs and therefore stability curves unless we fix $N({\rm H})$. 
 
Figure~\ref{fig_ther_curve1} shows the stability curves obtained for Obs~1-4 using model B. The vertical line indicates the $\log{(\xi)}$ obtained from the best-fit described in Section~\ref{sec_dis}. In all cases, the value of $\log{(\xi)}$ falls within a thermally stable region of the stability curve (defined by a positive slope in such a curve). Also, due to the low fraction of photons with energies >5$\times 10^{4}$ keV, we found no differences in the stability curves when using models C and D as ionizing SEDs. Middle and right lower panels, in the same Figure, show the stability curves for Obs~6 and 7 obtained using Model~C and using Model~D, respectively. For comparison, we also plot the stability curve of Obs~1. The shaded region corresponds to the uncertainties of the stability curves for heating and cooling errors of $15\%$ while the black stars indicate the best fit parameters obtained for Obs~1 (see Section~\ref{sec_disppaer} for an explanation about the horizontal arrows). When using model D the stability curve is practically identical between Obs~1-4 and~6 while for Obs~7, it evolves to a higher Compton temperature and with more unstable regions, characteristic of harder SEDs. When using model C, the curve shows a lower Compton temperature and more stable regions for Obs~6, compared to Obs~1-4, but this is likely a consequence of the degeneracies in the $\Gamma$-$N({\rm H})$ parameters of the model (see Figure~\ref{fig_contour}).   

 Finally, we have considered the effect of dust scattering in the spectra of 4U~1630-47, which has been reported previously  \citep{hor14,nei14,kal18}. The overall effect of the scattering in the spectra (and therefore in the SEDs) consist of the hardening of the continuum due to the $E^{-2}$ dependence of the dust model. Such effect could be larger if dust is close to the observer. For the current observations, we have found that the inclusion of a dust scattering component (namely {\tt xscat} in the {\sc xspec} software) does not improve the statistic of the fits, i.e. the dust contribution cannot be determined with our observations, which are degenerate to models including or not dust scattering. However, despite the model degeneracy, the dust scattering will undoubtedly have an effect in the recovered SED. Therefore, we tested its effect in Obs~6 by performing fits including dust scattering in two extreme cases (i.e. by fixing the location of the dust to be very close and far away from the source) to compare with the model without dust scattering. In both cases, the inclusion of dust scattering makes the recovered SED softer and therefore contributes to further enforce our conclusion that the wind in 4U~1630-47 has disappeared before the SED hardens enough to cause a thermal instability.             
 
 \section{Discussion}\label{sec_dis2}
We have performed a detailed analysis of six high-resolution {\it Chandra} spectra of the LMXB 4U~1630-47. We discuss below the findings of
our study. 
 
\subsection{Launching mechanism}\label{sec_launching}
From Obs~1-4, we have estimated the launching radius of the wind to be at $1.3$--$2.1$ $\times10^{11}$~cm, although, as described in Section~\ref{sec_dis}, such values depend on the hydrogen plasma density. It has been shown that a thermally driven wind can be launched at $\sim 10^{10} -5\times 10^{12}$ cm  \citep{beg83,woo96,hig17}. Radiative pressure due to line transitions in the UV energy band is negligible in this case due to the high disc temperature, which causes the SEDs to peak in the X-ray energy band, requiring unphysically large X-ray opacities to allow the line force to launch the wind \citep{pro02}, although \citet{wat18} point to the possibility that line radiation pressure is also present in X-ray binaries. We suggest that thermal pressure is the dominant process, as was proposed by previous analyses of the source \citep{kub07,dia14}, and given that velocities up to $\sim 600$ km/s can be reproduced with such a mechanism \citep[see e.g.][]{hig15}.  In principle, a magnetically driven wind cannot be excluded \citep{kub07}. However, more simulations are needed to understand if the parameters of the observed wind can be reproduced by a magnetically driven wind \citep[see e.g.][for cases where it may not]{wat18}.

It is important to note that the launching radius obtained from the ionization parameter depends on the SED and therefore on the continuum modeling.  In addition, systematic uncertainties due to differences in the calibration of the instruments need to be taken into account when computing fluxes. \citet{mad17} compute cross-normalization constants between {\it Chandra}, {\it Suzaku}, {\it NuSTAR}, {\it Swift} and {\it XMM-Newton} instruments in the 1--3 keV and 5--7 keV energy ranges using the quasar 3C~273 and the BL Lac PKS~2155-30 as calibration sources. They found that differences in flux measurements between the instruments in these energy ranges can be up to $\sim 15\%$. Also, the model degeneracy described in Section~\ref{sec_cont} increases the uncertainties in the flux calculation. For example, there is a factor 5 difference in the predicted 0.01-13 keV flux between Models C and D for Obs~6. 

The electron density value assumed affects the launching radius estimation as well. For example, if we include the instrumental flux uncertainty ($\sim 15\%$) and consider density values of $10^{11}$--$10^{13}$ cm$^{-3}$ we can estimate launching radii varying in the range $0.1$--$3.6$ $\times10^{11}$~cm for Obs~1--4, which then includes the values previously found by other authors.
 
In conclusion, because any outcome about the plasma physical conditions, such as the wind launching radius (and hence the analysis of the launching mechanism), depends on the flux obtained from the continuum fitting and on the hardness of the spectrum via the value of $\xi$, it is crucial (1) to construct the most accurate SED by including all information available from multiple energy bands  (e.g. we included {\it Swift/BAT} measurements as a fitting constraint) and (2) to consider the uncertainties inherent to the instruments.

 \subsection{The disappearance of the wind}\label{sec_disppaer}
 
Given that disc winds are mainly found in soft states of BH LMXBs while radio jets are found mainly in hard states it has been suggested that disc winds and jets are mutually exclusive \citep{neil09}. Although recent studies have shown the presence of disc wind signatures in hard accretion states \citep{hom16,mun17,all18}, for 4U~1630-47 it is clear that the disc wind has already disappeared in Obs~6, well before reaching the hard state. Among the theories to explain the absence of a wind in the hard-state are: thermal instabilities \citep{cha13,bia17}, full ionization of the plasma \citep{ued10,dia12,dia14,dia16} and disc geometry changes \citep{ued10,mil12b,pon12}. However, and given that the wind has already disappeared during Obs~6, that we classify as a relatively soft-intermediate state, and Obs~7 (see Figure~\ref{limit_nh}), we next checked if any of the above reasons could be used to explain the absence of the wind.

Thermal instabilities have been proposed as explanation for the disappearance of the wind during the hard state \citep{cha13,bia17}. Following \citet{cha16} and \citet{bia17}, if we assume that the the physical properties of the plasma do not change in the transition between different accretion states (i.e. $nr^{2}=$constant), we can trace the path from the best-fit value obtained in Obs~1 to the stability curves obtained for Obs~6 and 7.  These transitions are indicated in Figure~\ref{fig_ther_curve1}  by horizontal arrows.  We found that, under this assumption, the best-fits lie in a thermally stable region of the curve of stability when reaching Obs~6 and therefore the absence of absorption lines related to the wind is not expected due to the thermal equilibrium state of the gas predicted. In the case of Obs~7 the stability curves indicate that the best-fit parameters lie in a slightly unstable region. 

Next, we checked whether the plasma could be fully ionized in Obs~6 and~7 and therefore not visible via line absorption. Figure~\ref{fig_ion_fractions} shows the ion fractions for {\rm Fe}~{\sc XXIV}, {\rm Fe}~{\sc XXV} and {\rm Fe}~{\sc XXVI} as function of $\log{(\xi)}$ for the Obs~1 (solid lines), Obs~6 (dotted lines) and Obs~7 (dashed lines). It is clear from the plot that {\rm Fe}~{\sc XXV} and {\rm Fe}~{\sc XXVI} absorption lines should be observed for Obs~6 given the predicted ion ratios for the $\log{(\xi)}$ obtained  when assuming the  $nr^{2}$=constant condition in the transition between observations (dotted vertical line). Also, a small fraction of {\rm Fe}~{\sc XXVI} should be observed for Obs~7, although such fraction is lower when considering model C and D.

This is, thermal instabilities or over-ionization cannot explain the absence of absorption lines associated to the disc wind in Obs~6 and Obs~7. However, we note that the assumption of $nr^{2}$=constant is not correct for thermal winds \citep{don18}. In particular, \citet{don18} show that during hard states the Compton temperature is higher and consequently winds could be launched from radii that are smaller by up to one order of magnitude with respect to soft states. Indeed, when using model D, the Compton temperature for Obs~7 is significantly higher than for Obs~1-4 (see Figure~\ref{fig_ther_curve1}). In contrast, the assumption of  $nr^{2}$=constant might be correct for our Obs~6, where a large change in the Compton temperature is not observed with the same model. 
   
Therefore, we went on to try to find an alternative explanation for the absence of the wind in Obs~6. \citet{dyd17} showed that the thermal equilibrium branches of the stability curve might never be reached if insufficient flux is heating the flow. High flux cases, on the other hand, can lead to acceleration of the flow. Such acceleration produces a decrease in the column density of the plasma at given $\xi$ and therefore the absorption lines will be not detected due to the low $N({\rm H})$ predicted. \citet{dyd17}, in particular, used  SEDs obtained from 4U~1630-47 Obs~6 and 7 as a template for different states of X-ray binaries. They found that for Obs~6 the absorption measure distribution, i.e. the distribution of the absorber column density with the ionization parameter along the line of sight, shows two regions with very low column densities, when adiabatic cooling becomes important. However, stable plasma regions are predicted by their model, one with $5\times 10^{22}$ cm$^{-2}$ $< N({\rm H}) < 1\times 10^{23}$ cm$^{-2}$ for an ionization parameter $2.0<\log{\xi} < 2.1$ and a second one with $1\times 10^{21}$ cm$^{-2}$ $< N({\rm H}) < 1\times 10^{22}$ cm$^{-2}$ for $3.1<\log{\xi} < 4.0$. Such lower limits for the column densities are higher than the upper limits obtained from our fits (see Figure~\ref{limit_nh}).
 
Given these results, we propose as an alternative possibility that the hot plasma has been exhausted during the soft state due to its continuous outflow and that ``new'' fresh plasma could not be sufficiently heated due to the lower temperature of the disc as the source starts the transition to the hard state. Finally, we found no indications that matter has been diverted from the wind into the jet but we cannot rule out the existence of a weak jet given the upper limits obtained  from our radio observations (see Section~\ref{sec_radio}).
   
\section{Conclusions and summary}\label{sec_con}
We have analyzed six {\it Chandra} high-resolution spectra of the LMXB~4U~1630-47 obtained during the transition between soft and hard accretion states.  We included {\it Swift/BAT} data in the 15-50~keV range as a constraint in the hard-energy range for the data fitting. We found that different phenomenological models can be used to fit the continuum. From the {\tt diskbb} component we estimated a disc inner radius between 33 km and 36 km (assuming a disc inclination of 75$^{\circ}$). Absorption lines are only identified for Obs~1-4, which correspond to a soft accretion state. Common lines include {\rm Fe}~{\sc XXVI} K$\alpha$, K$\beta$, {\rm Fe}~{\sc XXV} K$\alpha$, K$\beta$ and {\rm Ca}~{\sc XX} K$\alpha$ while {\rm Ar}~{\sc XVIII}, {\rm S}~{\sc XVI} and {\rm Si}~{\sc XIV} ions are only identified in Obs~3. We noted that the {\rm Fe}~{\sc XXV} K$\alpha$ line in Obs~3 may be blended with the {\rm Fe}~{\sc XXIV} K$\alpha$ line.

We used the {\tt warmabs} photoionization model to fit the spectra of Obs~1-4. The best-fit parameters are similar between the observations except for the decreasing of $\log{(\xi)}$ in Obs~3, which leads to the formation of absorption lines associated to ions in lower ionization states. We inferred launching radii between $(1.3\sim 2.0)\times 10^{11}$~cm and column densities $N({\rm H})> 10^{23}$ cm$^{-2}$. The launching radius indicates that that thermal pressure is likely to be the dominant launching mechanism for the wind. We pointed out that discrepancies between the fluxes obtained from the continuum fitting in comparison with previous analyses of the same source can be due to instrumental systematic uncertainties as well as uncertainties in the modeling.
 
We computed stability curves for all observations using the {\sc xstar} photoionization code. We found that the best-fit parameters obtained for Obs~1-4 lie in thermally stable parts of the curve. In the case of Obs~6 and 7 we found a solution thermally stable if we consider that $nr^{2}$=constant. In this sense, thermal instabilities cannot explain the absence of absorption lines associated to the disc wind in Obs~6, before reaching the hard state. From the radio observations we found no indications that the jet has diverted matter from the wind. The discrepancies between the observation and predictions from photoionization models may indicate an acceleration of the flow at the end of the soft state, producing a decrease in the plasma column density, or that the plasma has been exhausted during the soft state.  More observations of LMXB systems during transitional states will confirm the proposed scenarios.
 
 \section{Acknowledgements}
The National Radio Astronomy Observatory is a facility of the National Science Foundation operated under cooperative agreement by Associated Universities, Inc. EG acknowledge support by the DFG cluster of excellence `Origin and Structure of the Universe'. JCAM-J is the recipient of an Australian Research Council Future Fellowship (FT140101082).
            
\bibliographystyle{mnras}

\begin{thebibliography}{}
\makeatletter
\relax
\def\mn@urlcharsother{\let\do\@makeother \do\$\do\&\do\#\do\^\do\_\do\%\do\~}
\def\mn@doi{\begingroup\mn@urlcharsother \@ifnextchar [ {\mn@doi@}
  {\mn@doi@[]}}
\def\mn@doi@[#1]#2{\def\@tempa{#1}\ifx\@tempa\@empty \href
  {http://dx.doi.org/#2} {doi:#2}\else \href {http://dx.doi.org/#2} {#1}\fi
  \endgroup}
\def\mn@eprint#1#2{\mn@eprint@#1:#2::\@nil}
\def\mn@eprint@arXiv#1{\href {http://arxiv.org/abs/#1} {{\tt arXiv:#1}}}
\def\mn@eprint@dblp#1{\href {http://dblp.uni-trier.de/rec/bibtex/#1.xml}
  {dblp:#1}}
\def\mn@eprint@#1:#2:#3:#4\@nil{\def\@tempa {#1}\def\@tempb {#2}\def\@tempc
  {#3}\ifx \@tempc \@empty \let \@tempc \@tempb \let \@tempb \@tempa \fi \ifx
  \@tempb \@empty \def\@tempb {arXiv}\fi \@ifundefined
  {mn@eprint@\@tempb}{\@tempb:\@tempc}{\expandafter \expandafter \csname
  mn@eprint@\@tempb\endcsname \expandafter{\@tempc}}}

\bibitem[\protect\citeauthoryear{{Abe}, {Fukazawa}, {Kubota}, {Kasama}  \&
  {Makishima}}{{Abe} et~al.}{2005}]{abe05}
{Abe} Y.,  {Fukazawa} Y.,  {Kubota} A.,  {Kasama} D.,   {Makishima} K.,  2005,
  \mn@doi [\pasj] {10.1093/pasj/57.4.629}, \href
  {http://adsabs.harvard.edu/abs/2005PASJ...57..629A} {57, 629}

\bibitem[\protect\citeauthoryear{{Allen}, {Schulz}, {Homan}, {Neilsen}, {Nowak}
   \& {Chakrabarty}}{{Allen} et~al.}{2018}]{all18}
{Allen} J.~L.,  {Schulz} N.~S.,  {Homan} J.,  {Neilsen} J.,  {Nowak} M.~A.,
  {Chakrabarty} D.,  2018, \mn@doi [\apj] {10.3847/1538-4357/aac2d1}, \href
  {http://adsabs.harvard.edu/abs/2018ApJ...861...26A} {861, 26}

\bibitem[\protect\citeauthoryear{{Begelman}, {McKee}  \& {Shields}}{{Begelman}
  et~al.}{1983}]{beg83}
{Begelman} M.~C.,  {McKee} C.~F.,   {Shields} G.~A.,  1983, \mn@doi [\apj]
  {10.1086/161178}, \href {http://adsabs.harvard.edu/abs/1983ApJ...271...70B}
  {271, 70}

\bibitem[\protect\citeauthoryear{{Bianchi}, {Ponti}, {Mu{\~n}oz-Darias}  \&
  {Petrucci}}{{Bianchi} et~al.}{2017}]{bia17}
{Bianchi} S.,  {Ponti} G.,  {Mu{\~n}oz-Darias} T.,   {Petrucci} P.-O.,  2017,
  \mn@doi [\mnras] {10.1093/mnras/stx2173}, \href
  {http://adsabs.harvard.edu/abs/2017MNRAS.472.2454B} {472, 2454}

\bibitem[\protect\citeauthoryear{{Blandford} \& {Payne}}{{Blandford} \&
  {Payne}}{1982}]{bla82}
{Blandford} R.~D.,  {Payne} D.~G.,  1982, \mn@doi [\mnras]
  {10.1093/mnras/199.4.883}, \href
  {http://adsabs.harvard.edu/abs/1982MNRAS.199..883B} {199, 883}

\bibitem[\protect\citeauthoryear{{Cash}}{{Cash}}{1979}]{cas79}
{Cash} W.,  1979, \mn@doi [\apj] {10.1086/156922}, \href
  {http://adsabs.harvard.edu/abs/1979ApJ...228..939C} {228, 939}

\bibitem[\protect\citeauthoryear{{Chakravorty}, {Kembhavi}, {Elvis}  \&
  {Ferland}}{{Chakravorty} et~al.}{2009}]{cha09}
{Chakravorty} S.,  {Kembhavi} A.~K.,  {Elvis} M.,   {Ferland} G.,  2009,
  \mn@doi [\mnras] {10.1111/j.1365-2966.2008.14249.x}, \href
  {http://adsabs.harvard.edu/abs/2009MNRAS.393...83C} {393, 83}

\bibitem[\protect\citeauthoryear{{Chakravorty}, {Lee}  \&
  {Neilsen}}{{Chakravorty} et~al.}{2013}]{cha13}
{Chakravorty} S.,  {Lee} J.~C.,   {Neilsen} J.,  2013, \mn@doi [\mnras]
  {10.1093/mnras/stt1593}, \href
  {http://adsabs.harvard.edu/abs/2013MNRAS.436..560C} {436, 560}

\bibitem[\protect\citeauthoryear{{Chakravorty} et~al.,}{{Chakravorty}
  et~al.}{2016}]{cha16}
{Chakravorty} S.,  et~al., 2016, \mn@doi [\aap] {10.1051/0004-6361/201527163},
  \href {http://adsabs.harvard.edu/abs/2016A%26A...589A.119C} {589, A119}

\bibitem[\protect\citeauthoryear{{Churazov}, {Gilfanov}, {Forman}  \&
  {Jones}}{{Churazov} et~al.}{1996}]{chu96}
{Churazov} E.,  {Gilfanov} M.,  {Forman} W.,   {Jones} C.,  1996, \mn@doi
  [\apj] {10.1086/177997}, \href
  {http://adsabs.harvard.edu/abs/1996ApJ...471..673C} {471, 673}

\bibitem[\protect\citeauthoryear{{D{\'{\i}}az Trigo} \& {Boirin}}{{D{\'{\i}}az
  Trigo} \& {Boirin}}{2016}]{dia16}
{D{\'{\i}}az Trigo} M.,  {Boirin} L.,  2016, \mn@doi [Astronomische
  Nachrichten] {10.1002/asna.201612315}, \href
  {http://adsabs.harvard.edu/abs/2016AN....337..368D} {337, 368}

\bibitem[\protect\citeauthoryear{{D{\'{\i}}az Trigo}, {Parmar}, {Miller},
  {Kuulkers}  \& {Caballero-Garc{\'{\i}}a}}{{D{\'{\i}}az Trigo}
  et~al.}{2007}]{dia07}
{D{\'{\i}}az Trigo} M.,  {Parmar} A.~N.,  {Miller} J.,  {Kuulkers} E.,
  {Caballero-Garc{\'{\i}}a} M.~D.,  2007, \mn@doi [\aap]
  {10.1051/0004-6361:20065406}, \href
  {http://adsabs.harvard.edu/abs/2007A%26A...462..657D} {462, 657}

\bibitem[\protect\citeauthoryear{{D{\'{\i}}az Trigo}, {Sidoli}, {Boirin}  \&
  {Parmar}}{{D{\'{\i}}az Trigo} et~al.}{2012}]{dia12}
{D{\'{\i}}az Trigo} M.,  {Sidoli} L.,  {Boirin} L.,   {Parmar} A.~N.,  2012,
  \mn@doi [\aap] {10.1051/0004-6361/201219049}, \href
  {http://adsabs.harvard.edu/abs/2012A%26A...543A..50D} {543, A50}

\bibitem[\protect\citeauthoryear{{D{\'{\i}}az Trigo}, {Miller-Jones},
  {Migliari}, {Broderick}  \& {Tzioumis}}{{D{\'{\i}}az Trigo}
  et~al.}{2013}]{dia13}
{D{\'{\i}}az Trigo} M.,  {Miller-Jones} J.~C.~A.,  {Migliari} S.,  {Broderick}
  J.~W.,   {Tzioumis} T.,  2013, \mn@doi [\nat] {10.1038/nature12672}, \href
  {http://adsabs.harvard.edu/abs/2013Natur.504..260D} {504, 260}

\bibitem[\protect\citeauthoryear{{D{\'{\i}}az Trigo}, {Migliari},
  {Miller-Jones}  \& {Guainazzi}}{{D{\'{\i}}az Trigo} et~al.}{2014}]{dia14}
{D{\'{\i}}az Trigo} M.,  {Migliari} S.,  {Miller-Jones} J.~C.~A.,   {Guainazzi}
  M.,  2014, \mn@doi [\aap] {10.1051/0004-6361/201424554}, \href
  {http://adsabs.harvard.edu/abs/2014A%26A...571A..76D} {571, A76}

\bibitem[\protect\citeauthoryear{{Done}, {Tomaru}  \& {Takahashi}}{{Done}
  et~al.}{2018}]{don18}
{Done} C.,  {Tomaru} R.,   {Takahashi} T.,  2018, \mn@doi [\mnras]
  {10.1093/mnras/stx2400}, \href
  {http://adsabs.harvard.edu/abs/2018MNRAS.473..838D} {473, 838}

\bibitem[\protect\citeauthoryear{{Dyda}, {Dannen}, {Waters}  \& {Proga}}{{Dyda}
  et~al.}{2017}]{dyd17}
{Dyda} S.,  {Dannen} R.,  {Waters} T.,   {Proga} D.,  2017, \mn@doi [\mnras]
  {10.1093/mnras/stx406}, \href
  {http://adsabs.harvard.edu/abs/2017MNRAS.467.4161D} {467, 4161}

\bibitem[\protect\citeauthoryear{{Fender} \& {Belloni}}{{Fender} \&
  {Belloni}}{2012}]{fen12}
{Fender} R.,  {Belloni} T.,  2012, \mn@doi [Science] {10.1126/science.1221790},
  \href {http://adsabs.harvard.edu/abs/2012Sci...337..540F} {337, 540}

\bibitem[\protect\citeauthoryear{{Fender}, {Belloni}  \& {Gallo}}{{Fender}
  et~al.}{2004}]{fen04}
{Fender} R.~P.,  {Belloni} T.~M.,   {Gallo} E.,  2004, \mn@doi [\mnras]
  {10.1111/j.1365-2966.2004.08384.x}, \href
  {http://adsabs.harvard.edu/abs/2004MNRAS.355.1105F} {355, 1105}

\bibitem[\protect\citeauthoryear{{Fukumura}, {Kazanas}, {Shrader}, {Behar},
  {Tombesi}  \& {Contopoulos}}{{Fukumura} et~al.}{2017}]{fuk17}
{Fukumura} K.,  {Kazanas} D.,  {Shrader} C.,  {Behar} E.,  {Tombesi} F.,
  {Contopoulos} I.,  2017, \mn@doi [Nature Astronomy]
  {10.1038/s41550-017-0062}, \href
  {http://adsabs.harvard.edu/abs/2017NatAs...1E..62F} {1, 0062}

\bibitem[\protect\citeauthoryear{{Gatuzz} \& {Churazov}}{{Gatuzz} \&
  {Churazov}}{2018}]{gat18a}
{Gatuzz} E.,  {Churazov} E.,  2018, \mn@doi [\mnras] {10.1093/mnras/stx2776},
  \href {http://adsabs.harvard.edu/abs/2018MNRAS.474..696G} {474, 696}

\bibitem[\protect\citeauthoryear{{Grevesse} \& {Sauval}}{{Grevesse} \&
  {Sauval}}{1998}]{gre98}
{Grevesse} N.,  {Sauval} A.~J.,  1998, \mn@doi [\ssr]
  {10.1023/A:1005161325181}, \href
  {http://adsabs.harvard.edu/abs/1998SSRv...85..161G} {85, 161}

\bibitem[\protect\citeauthoryear{{Hanke}, {Wilms}, {Nowak}, {Pottschmidt},
  {Schulz}  \& {Lee}}{{Hanke} et~al.}{2009}]{han09}
{Hanke} M.,  {Wilms} J.,  {Nowak} M.~A.,  {Pottschmidt} K.,  {Schulz} N.~S.,
  {Lee} J.~C.,  2009, \mn@doi [\apj] {10.1088/0004-637X/690/1/330}, \href
  {http://adsabs.harvard.edu/abs/2009ApJ...690..330H} {690, 330}

\bibitem[\protect\citeauthoryear{{Hashizume}, {Ohsuga}, {Kawashima}  \&
  {Tanaka}}{{Hashizume} et~al.}{2015}]{has15}
{Hashizume} K.,  {Ohsuga} K.,  {Kawashima} T.,   {Tanaka} M.,  2015, \mn@doi
  [\pasj] {10.1093/pasj/psu132}, \href
  {http://adsabs.harvard.edu/abs/2015PASJ...67...58H} {67, 58}

\bibitem[\protect\citeauthoryear{{Higginbottom} \& {Proga}}{{Higginbottom} \&
  {Proga}}{2015}]{hig15}
{Higginbottom} N.,  {Proga} D.,  2015, \mn@doi [\apj]
  {10.1088/0004-637X/807/1/107}, \href
  {http://adsabs.harvard.edu/abs/2015ApJ...807..107H} {807, 107}

\bibitem[\protect\citeauthoryear{{Higginbottom}, {Proga}, {Knigge}  \&
  {Long}}{{Higginbottom} et~al.}{2017}]{hig17}
{Higginbottom} N.,  {Proga} D.,  {Knigge} C.,   {Long} K.~S.,  2017, \mn@doi
  [\apj] {10.3847/1538-4357/836/1/42}, \href
  {http://adsabs.harvard.edu/abs/2017ApJ...836...42H} {836, 42}

\bibitem[\protect\citeauthoryear{{Hjellming} et~al.,}{{Hjellming}
  et~al.}{1999}]{hje99}
{Hjellming} R.~M.,  et~al., 1999, \mn@doi [\apj] {10.1086/306948}, \href
  {http://adsabs.harvard.edu/abs/1999ApJ...514..383H} {514, 383}

\bibitem[\protect\citeauthoryear{{Homan}, {Neilsen}, {Allen}, {Chakrabarty},
  {Fender}, {Fridriksson}, {Remillard}  \& {Schulz}}{{Homan}
  et~al.}{2016}]{hom16}
{Homan} J.,  {Neilsen} J.,  {Allen} J.~L.,  {Chakrabarty} D.,  {Fender} R.,
  {Fridriksson} J.~K.,  {Remillard} R.~A.,   {Schulz} N.,  2016, \mn@doi
  [\apjl] {10.3847/2041-8205/830/1/L5}, \href
  {http://adsabs.harvard.edu/abs/2016ApJ...830L...5H} {830, L5}

\bibitem[\protect\citeauthoryear{{Hori} et~al.,}{{Hori} et~al.}{2014}]{hor14}
{Hori} T.,  et~al., 2014, \mn@doi [\apj] {10.1088/0004-637X/790/1/20}, \href
  {http://adsabs.harvard.edu/abs/2014ApJ...790...20H} {790, 20}

\bibitem[\protect\citeauthoryear{{Houck} \& {Denicola}}{{Houck} \&
  {Denicola}}{2000}]{hou00}
{Houck} J.~C.,  {Denicola} L.~A.,  2000, in {Manset} N.,  {Veillet} C.,
  {Crabtree} D.,  eds,  Astronomical Society of the Pacific Conference Series
  Vol. 216, Astronomical Data Analysis Software and Systems IX. p.~591

\bibitem[\protect\citeauthoryear{{Joachimi}, {Gatuzz}, {Garc{\'{\i}}a}  \&
  {Kallman}}{{Joachimi} et~al.}{2016}]{joa16}
{Joachimi} K.,  {Gatuzz} E.,  {Garc{\'{\i}}a} J.~A.,   {Kallman} T.~R.,  2016,
  \mn@doi [\mnras] {10.1093/mnras/stw1371}, \href
  {http://adsabs.harvard.edu/abs/2016MNRAS.461..352J} {461, 352}

\bibitem[\protect\citeauthoryear{{Jones}, {Forman}, {Tananbaum}  \&
  {Turner}}{{Jones} et~al.}{1976}]{jon76}
{Jones} C.,  {Forman} W.,  {Tananbaum} H.,   {Turner} M.~J.~L.,  1976, \mn@doi
  [\apjl] {10.1086/182291}, \href
  {http://adsabs.harvard.edu/abs/1976ApJ...210L...9J} {210, L9}

\bibitem[\protect\citeauthoryear{{Kalemci}, {Maccarone}  \&
  {Tomsick}}{{Kalemci} et~al.}{2018}]{kal18}
{Kalemci} E.,  {Maccarone} T.~J.,   {Tomsick} J.~A.,  2018, \mn@doi [\apj]
  {10.3847/1538-4357/aabcd3}, \href
  {http://adsabs.harvard.edu/abs/2018ApJ...859...88K} {859, 88}

\bibitem[\protect\citeauthoryear{{Kallman}, {Bautista}, {Goriely}, {Mendoza},
  {Miller}, {Palmeri}, {Quinet}  \& {Raymond}}{{Kallman} et~al.}{2009}]{kal09}
{Kallman} T.~R.,  {Bautista} M.~A.,  {Goriely} S.,  {Mendoza} C.,  {Miller}
  J.~M.,  {Palmeri} P.,  {Quinet} P.,   {Raymond} J.,  2009, \mn@doi [\apj]
  {10.1088/0004-637X/701/2/865}, \href
  {http://adsabs.harvard.edu/abs/2009ApJ...701..865K} {701, 865}

\bibitem[\protect\citeauthoryear{{King} et~al.,}{{King} et~al.}{2014}]{kin14}
{King} A.~L.,  et~al., 2014, \mn@doi [\apjl] {10.1088/2041-8205/784/1/L2},
  \href {http://adsabs.harvard.edu/abs/2014ApJ...784L...2K} {784, L2}

\bibitem[\protect\citeauthoryear{{Kotani}, {Ebisawa}, {Dotani}, {Inoue},
  {Nagase}, {Tanaka}  \& {Ueda}}{{Kotani} et~al.}{2000}]{kot00}
{Kotani} T.,  {Ebisawa} K.,  {Dotani} T.,  {Inoue} H.,  {Nagase} F.,  {Tanaka}
  Y.,   {Ueda} Y.,  2000, \mn@doi [\apj] {10.1086/309200}, \href
  {http://adsabs.harvard.edu/abs/2000ApJ...539..413K} {539, 413}

\bibitem[\protect\citeauthoryear{{Krolik}, {McKee}  \& {Tarter}}{{Krolik}
  et~al.}{1981}]{kro81}
{Krolik} J.~H.,  {McKee} C.~F.,   {Tarter} C.~B.,  1981, \mn@doi [\apj]
  {10.1086/159303}, \href {http://adsabs.harvard.edu/abs/1981ApJ...249..422K}
  {249, 422}

\bibitem[\protect\citeauthoryear{{Kubota} et~al.,}{{Kubota}
  et~al.}{2007}]{kub07}
{Kubota} A.,  et~al., 2007, \mn@doi [\pasj] {10.1093/pasj/59.sp1.S185}, \href
  {http://adsabs.harvard.edu/abs/2007PASJ...59S.185K} {59, 185}

\bibitem[\protect\citeauthoryear{{Kuulkers}, {Parmar}, {Kitamoto}, {Cominsky}
  \& {Sood}}{{Kuulkers} et~al.}{1997}]{kuu97}
{Kuulkers} E.,  {Parmar} A.~N.,  {Kitamoto} S.,  {Cominsky} L.~R.,   {Sood}
  R.~K.,  1997, \mn@doi [\mnras] {10.1093/mnras/291.1.81}, \href
  {http://adsabs.harvard.edu/abs/1997MNRAS.291...81K} {291, 81}

\bibitem[\protect\citeauthoryear{{Kuulkers}, {Wijnands}, {Belloni},
  {M{\'e}ndez}, {van der Klis}  \& {van Paradijs}}{{Kuulkers}
  et~al.}{1998}]{kuu98}
{Kuulkers} E.,  {Wijnands} R.,  {Belloni} T.,  {M{\'e}ndez} M.,  {van der Klis}
  M.,   {van Paradijs} J.,  1998, \mn@doi [\apj] {10.1086/305248}, \href
  {http://adsabs.harvard.edu/abs/1998ApJ...494..753K} {494, 753}

\bibitem[\protect\citeauthoryear{{Li} \& {Begelman}}{{Li} \&
  {Begelman}}{2014}]{li14}
{Li} S.-L.,  {Begelman} M.~C.,  2014, \mn@doi [\apj]
  {10.1088/0004-637X/786/1/6}, \href
  {http://adsabs.harvard.edu/abs/2014ApJ...786....6L} {786, 6}

\bibitem[\protect\citeauthoryear{{Lilley} \& {Palmer}}{{Lilley} \&
  {Palmer}}{1968}]{lil68}
{Lilley} A.~E.,  {Palmer} P.,  1968, \mn@doi [\apjs] {10.1086/190172}, \href
  {http://adsabs.harvard.edu/abs/1968ApJS...16..143L} {16, 143}

\bibitem[\protect\citeauthoryear{{Luketic}, {Proga}, {Kallman}, {Raymond}  \&
  {Miller}}{{Luketic} et~al.}{2010}]{luk10}
{Luketic} S.,  {Proga} D.,  {Kallman} T.~R.,  {Raymond} J.~C.,   {Miller}
  J.~M.,  2010, \mn@doi [\apj] {10.1088/0004-637X/719/1/515}, \href
  {http://adsabs.harvard.edu/abs/2010ApJ...719..515L} {719, 515}

\bibitem[\protect\citeauthoryear{{Madsen}, {Beardmore}, {Forster}, {Guainazzi},
  {Marshall}, {Miller}, {Page}  \& {Stuhlinger}}{{Madsen} et~al.}{2017}]{mad17}
{Madsen} K.~K.,  {Beardmore} A.~P.,  {Forster} K.,  {Guainazzi} M.,  {Marshall}
  H.~L.,  {Miller} E.~D.,  {Page} K.~L.,   {Stuhlinger} M.,  2017, \mn@doi
  [\aj] {10.3847/1538-3881/153/1/2}, \href
  {http://adsabs.harvard.edu/abs/2017AJ....153....2M} {153, 2}

\bibitem[\protect\citeauthoryear{{Makishima}, {Maejima}, {Mitsuda}, {Bradt},
  {Remillard}, {Tuohy}, {Hoshi}  \& {Nakagawa}}{{Makishima}
  et~al.}{1986}]{mak86}
{Makishima} K.,  {Maejima} Y.,  {Mitsuda} K.,  {Bradt} H.~V.,  {Remillard}
  R.~A.,  {Tuohy} I.~R.,  {Hoshi} R.,   {Nakagawa} M.,  1986, \mn@doi [\apj]
  {10.1086/164534}, \href {http://adsabs.harvard.edu/abs/1986ApJ...308..635M}
  {308, 635}

\bibitem[\protect\citeauthoryear{{McClintock} \& {Remillard}}{{McClintock} \&
  {Remillard}}{2006}]{mcc06}
{McClintock} J.~E.,  {Remillard} R.~A.,  2006, {Black hole binaries}.
pp 157--213

\bibitem[\protect\citeauthoryear{{Medvedev}, {Khabibullin}, {Sazonov},
  {Churazov}  \& {Tsygankov}}{{Medvedev} et~al.}{2018}]{med18}
{Medvedev} P.~S.,  {Khabibullin} I.~I.,  {Sazonov} S.~Y.,  {Churazov} E.~M.,
  {Tsygankov} S.~S.,  2018, \mn@doi [Astronomy Letters]
  {10.1134/S1063773718060038}, \href
  {http://adsabs.harvard.edu/abs/2018AstL...44..390M} {44, 390}

\bibitem[\protect\citeauthoryear{{Miller}, {Raymond}, {Fabian}, {Steeghs},
  {Homan}, {Reynolds}, {van der Klis}  \& {Wijnands}}{{Miller}
  et~al.}{2006a}]{mil06a}
{Miller} J.~M.,  {Raymond} J.,  {Fabian} A.,  {Steeghs} D.,  {Homan} J.,
  {Reynolds} C.,  {van der Klis} M.,   {Wijnands} R.,  2006a, \mn@doi [\nat]
  {10.1038/nature04912}, \href
  {http://adsabs.harvard.edu/abs/2006Natur.441..953M} {441, 953}

\bibitem[\protect\citeauthoryear{{Miller} et~al.,}{{Miller}
  et~al.}{2006b}]{mil06dd}
{Miller} J.~M.,  et~al., 2006b, \mn@doi [\apj] {10.1086/504673}, \href
  {http://adsabs.harvard.edu/abs/2006ApJ...646..394M} {646, 394}

\bibitem[\protect\citeauthoryear{{Miller}, {Raymond}, {Reynolds}, {Fabian},
  {Kallman}  \& {Homan}}{{Miller} et~al.}{2008}]{mil08}
{Miller} J.~M.,  {Raymond} J.,  {Reynolds} C.~S.,  {Fabian} A.~C.,  {Kallman}
  T.~R.,   {Homan} J.,  2008, \mn@doi [\apj] {10.1086/588521}, \href
  {http://adsabs.harvard.edu/abs/2008ApJ...680.1359M} {680, 1359}

\bibitem[\protect\citeauthoryear{{Miller} et~al.,}{{Miller}
  et~al.}{2012}]{mil12b}
{Miller} J.~M.,  et~al., 2012, \mn@doi [\apjl] {10.1088/2041-8205/759/1/L6},
  \href {http://adsabs.harvard.edu/abs/2012ApJ...759L...6M} {759, L6}

\bibitem[\protect\citeauthoryear{{Miller} et~al.,}{{Miller}
  et~al.}{2016a}]{mil16c}
{Miller} J.~M.,  et~al., 2016a, \mn@doi [\apjl] {10.3847/2041-8205/821/1/L9},
  \href {http://adsabs.harvard.edu/abs/2016ApJ...821L...9M} {821, L9}

\bibitem[\protect\citeauthoryear{{Miller}, {Raymond}, {Cackett}, {Grinberg}  \&
  {Nowak}}{{Miller} et~al.}{2016b}]{mil16d}
{Miller} J.~M.,  {Raymond} J.,  {Cackett} E.,  {Grinberg} V.,   {Nowak} M.,
  2016b, \mn@doi [\apjl] {10.3847/2041-8205/822/1/L18}, \href
  {http://adsabs.harvard.edu/abs/2016ApJ...822L..18M} {822, L18}

\bibitem[\protect\citeauthoryear{{Mitsuda} et~al.,}{{Mitsuda}
  et~al.}{1984}]{mit84}
{Mitsuda} K.,  et~al., 1984, \pasj, \href
  {http://adsabs.harvard.edu/abs/1984PASJ...36..741M} {36, 741}

\bibitem[\protect\citeauthoryear{{Mu{\~n}oz-Darias} et~al.,}{{Mu{\~n}oz-Darias}
  et~al.}{2017}]{mun17}
{Mu{\~n}oz-Darias} T.,  et~al., 2017, \mn@doi [\mnras] {10.1093/mnrasl/slw222},
  \href {http://adsabs.harvard.edu/abs/2017MNRAS.465L.124M} {465, L124}

\bibitem[\protect\citeauthoryear{{Neilsen} \& {Homan}}{{Neilsen} \&
  {Homan}}{2012}]{nei12}
{Neilsen} J.,  {Homan} J.,  2012, \mn@doi [\apj] {10.1088/0004-637X/750/1/27},
  \href {http://adsabs.harvard.edu/abs/2012ApJ...750...27N} {750, 27}

\bibitem[\protect\citeauthoryear{{Neilsen} \& {Lee}}{{Neilsen} \&
  {Lee}}{2009a}]{nei09b}
{Neilsen} J.,  {Lee} J.~C.,  2009a, \mn@doi [\nat] {10.1038/nature07680}, \href
  {http://adsabs.harvard.edu/abs/2009Natur.458..481N} {458, 481}

\bibitem[\protect\citeauthoryear{{Neilsen} \& {Lee}}{{Neilsen} \&
  {Lee}}{2009b}]{neil09}
{Neilsen} J.,  {Lee} J.~C.,  2009b, \mn@doi [\nat] {10.1038/nature07680}, \href
  {http://adsabs.harvard.edu/abs/2009Natur.458..481N} {458, 481}

\bibitem[\protect\citeauthoryear{{Neilsen}, {Coriat}, {Fender}, {Lee}, {Ponti},
  {Tzioumis}, {Edwards}  \& {Broderick}}{{Neilsen} et~al.}{2014}]{nei14}
{Neilsen} J.,  {Coriat} M.,  {Fender} R.,  {Lee} J.~C.,  {Ponti} G.,
  {Tzioumis} A.~K.,  {Edwards} P.~G.,   {Broderick} J.~W.,  2014, \mn@doi
  [\apjl] {10.1088/2041-8205/784/1/L5}, \href
  {http://adsabs.harvard.edu/abs/2014ApJ...784L...5N} {784, L5}

\bibitem[\protect\citeauthoryear{{Netzer}}{{Netzer}}{2006}]{net06}
{Netzer} H.,  2006, \mn@doi [\apjl] {10.1086/510067}, \href
  {http://adsabs.harvard.edu/abs/2006ApJ...652L.117N} {652, L117}

\bibitem[\protect\citeauthoryear{{Parmar}, {Angelini}  \& {White}}{{Parmar}
  et~al.}{1995}]{par95}
{Parmar} A.~N.,  {Angelini} L.,   {White} N.~E.,  1995, \mn@doi [\apjl]
  {10.1086/309730}, \href {http://adsabs.harvard.edu/abs/1995ApJ...452L.129P}
  {452, L129}

\bibitem[\protect\citeauthoryear{{Ponti}, {Fender}, {Begelman}, {Dunn},
  {Neilsen}  \& {Coriat}}{{Ponti} et~al.}{2012}]{pon12}
{Ponti} G.,  {Fender} R.~P.,  {Begelman} M.~C.,  {Dunn} R.~J.~H.,  {Neilsen}
  J.,   {Coriat} M.,  2012, \mn@doi [\mnras]
  {10.1111/j.1745-3933.2012.01224.x}, \href
  {http://esoads.eso.org/abs/2012MNRAS.422L..11P} {422, L11}

\bibitem[\protect\citeauthoryear{{Proga}}{{Proga}}{2003}]{pro03}
{Proga} D.,  2003, \mn@doi [\apj] {10.1086/345897}, \href
  {http://adsabs.harvard.edu/abs/2003ApJ...585..406P} {585, 406}

\bibitem[\protect\citeauthoryear{{Proga} \& {Kallman}}{{Proga} \&
  {Kallman}}{2002}]{pro02}
{Proga} D.,  {Kallman} T.~R.,  2002, \mn@doi [\apj] {10.1086/324534}, \href
  {http://adsabs.harvard.edu/abs/2002ApJ...565..455P} {565, 455}

\bibitem[\protect\citeauthoryear{{Rahoui}, {Coriat}  \& {Lee}}{{Rahoui}
  et~al.}{2014}]{rah14}
{Rahoui} F.,  {Coriat} M.,   {Lee} J.~C.,  2014, \mn@doi [\mnras]
  {10.1093/mnras/stu977}, \href
  {http://adsabs.harvard.edu/abs/2014MNRAS.442.1610R} {442, 1610}

\bibitem[\protect\citeauthoryear{{R{\'o}{\.z}a{\'n}ska}, {Goosmann}, {Dumont}
  \& {Czerny}}{{R{\'o}{\.z}a{\'n}ska} et~al.}{2006}]{roz06}
{R{\'o}{\.z}a{\'n}ska} A.,  {Goosmann} R.,  {Dumont} A.-M.,   {Czerny} B.,
  2006, \mn@doi [\aap] {10.1051/0004-6361:20052723}, \href
  {http://adsabs.harvard.edu/abs/2006A%26A...452....1R} {452, 1}

\bibitem[\protect\citeauthoryear{{R{\'o}{\.z}a{\'n}ska}, {Madej},
  {Bagi{\'n}ska}, {Hryniewicz}  \& {Handzlik}}{{R{\'o}{\.z}a{\'n}ska}
  et~al.}{2014}]{roz14}
{R{\'o}{\.z}a{\'n}ska} A.,  {Madej} J.,  {Bagi{\'n}ska} P.,  {Hryniewicz} K.,
  {Handzlik} B.,  2014, \mn@doi [\aap] {10.1051/0004-6361/201321567}, \href
  {http://adsabs.harvard.edu/abs/2014A%26A...562A..81R} {562, A81}

\bibitem[\protect\citeauthoryear{{Seifina}, {Titarchuk}  \&
  {Shaposhnikov}}{{Seifina} et~al.}{2014}]{sei14}
{Seifina} E.,  {Titarchuk} L.,   {Shaposhnikov} N.,  2014, \mn@doi [\apj]
  {10.1088/0004-637X/789/1/57}, \href
  {http://adsabs.harvard.edu/abs/2014ApJ...789...57S} {789, 57}

\bibitem[\protect\citeauthoryear{{Shidatsu}, {Done}  \& {Ueda}}{{Shidatsu}
  et~al.}{2016}]{shi16}
{Shidatsu} M.,  {Done} C.,   {Ueda} Y.,  2016, \mn@doi [\apj]
  {10.3847/0004-637X/823/2/159}, \href
  {http://adsabs.harvard.edu/abs/2016ApJ...823..159S} {823, 159}

\bibitem[\protect\citeauthoryear{{Steiner}, {Narayan}, {McClintock}  \&
  {Ebisawa}}{{Steiner} et~al.}{2009}]{ste09}
{Steiner} J.~F.,  {Narayan} R.,  {McClintock} J.~E.,   {Ebisawa} K.,  2009,
  \mn@doi [\pasp] {10.1086/648535}, \href
  {http://adsabs.harvard.edu/abs/2009PASP..121.1279S} {121, 1279}

\bibitem[\protect\citeauthoryear{{Tarter}, {Tucker}  \& {Salpeter}}{{Tarter}
  et~al.}{1969}]{tar69}
{Tarter} C.~B.,  {Tucker} W.~H.,   {Salpeter} E.~E.,  1969, \mn@doi [\apj]
  {10.1086/150026}, \href {http://adsabs.harvard.edu/abs/1969ApJ...156..943T}
  {156, 943}

\bibitem[\protect\citeauthoryear{{Tetarenko}, {Lasota}, {Heinke}, {Dubus}  \&
  {Sivakoff}}{{Tetarenko} et~al.}{2018}]{tet18}
{Tetarenko} B.~E.,  {Lasota} J.-P.,  {Heinke} C.~O.,  {Dubus} G.,   {Sivakoff}
  G.~R.,  2018, \mn@doi [\nat] {10.1038/nature25159}, \href
  {http://adsabs.harvard.edu/abs/2018Natur.554...69T} {554, 69}

\bibitem[\protect\citeauthoryear{{Tofflemire}, {Orio}, {Page}, {Osborne},
  {Ciroi}, {Cracco}, {Di Mille}  \& {Maxwell}}{{Tofflemire}
  et~al.}{2013}]{tof13}
{Tofflemire} B.~M.,  {Orio} M.,  {Page} K.~L.,  {Osborne} J.~P.,  {Ciroi} S.,
  {Cracco} V.,  {Di Mille} F.,   {Maxwell} M.,  2013, \mn@doi [\apj]
  {10.1088/0004-637X/779/1/22}, \href
  {http://adsabs.harvard.edu/abs/2013ApJ...779...22T} {779, 22}

\bibitem[\protect\citeauthoryear{{Tomaru}, {Done}, {Odaka}, {Watanabe}  \&
  {Takahashi}}{{Tomaru} et~al.}{2018}]{tom18}
{Tomaru} R.,  {Done} C.,  {Odaka} H.,  {Watanabe} S.,   {Takahashi} T.,  2018,
  \mn@doi [\mnras] {10.1093/mnras/sty336}, \href
  {http://adsabs.harvard.edu/abs/2018MNRAS.476.1776T} {476, 1776}

\bibitem[\protect\citeauthoryear{{Tomsick}, {Lapshov}  \& {Kaaret}}{{Tomsick}
  et~al.}{1998}]{tom98}
{Tomsick} J.~A.,  {Lapshov} I.,   {Kaaret} P.,  1998, \mn@doi [\apj]
  {10.1086/305240}, \href {http://adsabs.harvard.edu/abs/1998ApJ...494..747T}
  {494, 747}

\bibitem[\protect\citeauthoryear{{Tomsick}, {Corbel}, {Goldwurm}  \&
  {Kaaret}}{{Tomsick} et~al.}{2005}]{tom05}
{Tomsick} J.~A.,  {Corbel} S.,  {Goldwurm} A.,   {Kaaret} P.,  2005, \mn@doi
  [\apj] {10.1086/431896}, \href
  {http://adsabs.harvard.edu/abs/2005ApJ...630..413T} {630, 413}

\bibitem[\protect\citeauthoryear{{Ueda}, {Yamaoka}  \& {Remillard}}{{Ueda}
  et~al.}{2009}]{ued09bb}
{Ueda} Y.,  {Yamaoka} K.,   {Remillard} R.,  2009, \mn@doi [\apj]
  {10.1088/0004-637X/695/2/888}, \href
  {http://adsabs.harvard.edu/abs/2009ApJ...695..888U} {695, 888}

\bibitem[\protect\citeauthoryear{{Ueda} et~al.,}{{Ueda} et~al.}{2010}]{ued10}
{Ueda} Y.,  et~al., 2010, \mn@doi [\apj] {10.1088/0004-637X/713/1/257}, \href
  {http://adsabs.harvard.edu/abs/2010ApJ...713..257U} {713, 257}

\bibitem[\protect\citeauthoryear{{Waters} \& {Proga}}{{Waters} \&
  {Proga}}{2018}]{wat18}
{Waters} T.,  {Proga} D.,  2018, preprint, \href
  {http://adsabs.harvard.edu/abs/2018arXiv180607322W} {} (\mn@eprint {arXiv}
  {1806.07322})

\bibitem[\protect\citeauthoryear{{Wilms}, {Allen}  \& {McCray}}{{Wilms}
  et~al.}{2000}]{wil00}
{Wilms} J.,  {Allen} A.,   {McCray} R.,  2000, \mn@doi [\apj] {10.1086/317016},
  \href {http://adsabs.harvard.edu/abs/2000ApJ...542..914W} {542, 914}

\bibitem[\protect\citeauthoryear{{Woods}, {Klein}, {Castor}, {McKee}  \&
  {Bell}}{{Woods} et~al.}{1996}]{woo96}
{Woods} D.~T.,  {Klein} R.~I.,  {Castor} J.~I.,  {McKee} C.~F.,   {Bell} J.~B.,
   1996, \mn@doi [\apj] {10.1086/177101}, \href
  {http://adsabs.harvard.edu/abs/1996ApJ...461..767W} {461, 767}

\makeatother
\end{thebibliography}

\end{document}